\documentclass[a4paper,11pt]{article}

\usepackage{jinstpub} 

\usepackage{color}

\usepackage{lineno}

\title{Muon veto system for the CROSS double-beta decay search experiment} 

\author[a]{A.S.~Barabash,}
\author[b]{L.~Berg{\'e},}
\author[b]{M.~Buchynska,}
\author[c,d]{J.M.~Calvo-Mozota,}
\author[e]{A.~Candela,}
\author[f,g]{P.~Carniti,}
\author[b]{M.~Chapellier,}
\author[b,1,2]{D.~Cintas,\note{Corresponding author.}\note{Now at IRFU, CEA, Universit\'e Paris-Saclay, Saclay, France.}}
\author[e]{A.~Corsi,}
\author[h]{I.~Dafinei,}
\author[i,j]{F.A.~Danevich,}
\author[e,1]{M.~De Deo,}
\author[b]{L.~Dumoulin,}
\author[k]{F.~Ferri,}
\author[b]{A.~Giuliani,}
\author[f]{C.~Gotti,}
\author[k]{P.~Gras,}
\author[e]{A.~Ianni,}
\author[i]{V.V.~Kobychev,}
\author[a]{S.I.~Konovalov,}
\author[b]{P.~Loaiza,}
\author[b]{P.~de~Marcillac,}
\author[b]{S.~Marnieros,}
\author[b]{C.A.~Marrache-Kikuchi,}
\author[l,m]{M.~Martinez,}
\author[k]{C.~Nones,}
\author[b]{E.~Olivieri,}
\author[l]{A.~Ortiz de Sol\'orzano,}
\author[b,c]{V.~Perez,}
\author[f]{G.~Pessina,}
\author[b]{D.V.~Poda,}
\author[e]{B.~Romualdi,}
\author[b]{Ph.~Rosier,}
\author[b]{R.~Serino,}
\author[i,j]{V.I.~Tretyak,}
\author[a]{V.I.~Umatov,}
\author[e]{G.~Ursini,}
\author[i]{M.M.~Zarytskyy,}
\author[k]{and A.~Zolotarova}

\affiliation[a]{National Research Center Kurchatov Institute, Kurchatov Complex of Theoretical and Experimental Physics, 117218 Moscow, Russia}
\affiliation[b]{Universit\'e Paris-Saclay, CNRS/IN2P3, IJCLab, 91405 Orsay, France}
\affiliation[c]{Laboratorio Subterr\'aneo de Canfranc, 22880 Canfranc-Estaci\'on, Spain}
\affiliation[d]{Escuela Superior de Ingenier\'ia y Tecnolog\'ia, Universidad Internacional de La Rioja, 26006 Logro\~no, Spain}
\affiliation[e]{INFN Laboratori Nazionali del Gran Sasso, I-67100 Assergi (AQ), Italy}
\affiliation[f]{INFN Sezione di Milano-Bicocca, I-20126 Milan, Italy}
\affiliation[g]{Universit\`{a} di Milano-Bicocca, Milano I-20126, Italy} 
\affiliation[h]{INFN Sezione di Roma, I-00185 Rome, Italy}
\affiliation[i]{Institute for Nuclear Research of NASU, 03028 Kyiv, Ukraine}
\affiliation[j]{Institute of Experimental and Applied Physics, CTU Prague, Prague, CA-11000, Czech Republic}
\affiliation[k]{IRFU, CEA, Université Paris-Saclay, 91191 Saclay, France}
\affiliation[l]{Centro de Astropart\'iculas y F\'isica de Altas Energ\'ias, Universidad de Zaragoza, 50009 Zaragoza, Spain}

\emailAdd{david.cintas@ijclab.in2p3.fr}
\emailAdd{massimiliano.dedeo@lngs.infn.it}

\abstract{  
In preparation to the CROSS experiment at the Canfranc underground laboratory (Spain) --- aiming to search for neutrinoless double-beta ($0\nu\beta\beta$) decay of $^{100}$Mo using low-temperature detectors with heat-scintillation readout --- we report on development of a dedicated muon veto system. The need for the muon veto in CROSS is caused by a comparatively high residual cosmic muon flux at the experimental site ($\sim$20 $\mu$/m$^2$/h), being a dominant background in the region of interest (ROI) at $\sim$3 MeV. Thus, we installed the muon veto system around the CROSS low-background setup, forming four lateral,  one top, and four bottom sectors. In this paper we describe the design, construction and operation of the CROSS muon veto system, as well as its optimization and validation by comparing  dedicated Monte Carlo (MC) simulations of muons with low-temperature measurements in the setup. 
We demonstrate a stable operation of the veto system with the average trigger rates compatible with MC simulations.  Also, we investigated two muon trigger logics based on coincidences with either 2 sectors or a single sector of the veto. 
The MC study shows that, in combination with the multiplicity cut of thermal detectors, these trigger logics allow to reject 99.2\% and 99.7\% of muon-induced events in the ROI, respectively. 
Despite a comparatively high dead time ($\sim$18\%) introduced by coincidences with any of nine sectors of the veto --- the adopted strategy --- the muon-induced background in the ROI of the CROSS experiment can be reduced down to $\sim$2 $\times 10^{-3}$ cnts/keV/kg/yr, i.e., an acceptable level compatible with a high-sensitivity $0\nu\beta\beta$ decay search foreseen in CROSS. 
}


\begin{document}
\maketitle
\flushbottom

\section{Introduction}
\label{sec:intro} 

The CROSS (Cryogenic Rare event Observatory with Surface Sensitivity) Collaboration is preparing a high-sensitivity cryogenic experiment to search for neutrinoless double-beta ($0\nu\beta\beta$) decay in $^{100}$Mo \cite{Bandac:2020} -- a hypothetical lepton-number-violating process, detecting of which can shed light on neutrino properties \cite{GomezCadenas:2023,Agostini:2023}. 
The CROSS experiment relies on a mature technology of thermal detectors, offering a high detection efficiency, a high energy resolution, and a scalability to a ton-scale array, as proved by the CUORE experiment \cite{Adams:2022a} ongoing in the Gran Sasso underground laboratory (LNGS, Italy).

The CROSS detector is constructed from 42 cryogenic calorimeters based on 36 lithium molybdate (Li$_2$MoO$_4$, LMO) crystals, 32 of them are $^{100}$Mo-enriched, 
and 6 $^{130}$Te-enriched tellurium dioxide (TeO$_2$, TeO) samples. 
The TeO array contain the same isotope of interest as in CUORE, but a factor 100 lower mass, which is not enough to do a competitive $0\nu\beta\beta$ decay search, but sufficient to demonstrate technology of enriched TeOs for future cryogenic experiments. 
Each crystal in CROSS is accompanied with a thin, wafer-type, cryogenic light detector, thus a single module has a dual heat-scintillation readout that allows particle identification, as demonstrated by CUPID-0 \cite{Azzolini:2022}, CUPID-Mo \cite{Augier:2022}, AMoRE-pilot \cite{Alenkov:2019jis}, and AMoRE-I \cite{Agrawal:2025amoreI} $0\nu\beta\beta$ decay search experiments exploiting this technique. The scintillation from LMOs (the Cherenkov radiation dominated signals from TeOs) in CROSS will be detected using a thin Ge or Si cryogenic light detector coupled to each crystal. The CROSS detector array has been assembled in a clean room of the IJCLab (Orsay, France) using a specially developed mechanical structure with a low amount of passive materials \cite{CROSSdetectorStructure:2024} --- a copper frame and supporting elements based on polylactide (PLA) that contribute 6\% and 0.04\% of the LMO mass, respectively. The CROSS detector has been safely transported to the Canfranc underground laboratory (LSC, Spain) and will soon be commissioned in a low-background cryogenic setup, already operational over 6 years \cite{Olivieri:2020,Armatol:2021b,CrossCupidTower:2023a,CROSSdetectorStructure:2024}.

A deep underground laboratory \cite{Ianni:2017} is a typical place for $0\nu\beta\beta$ decay experiments, as it allows a significant reduction in the flux of cosmic muons, characterized by a high penetration depth and a high energy, and thus a notable suppression of a muon-induced background in the region of interest (ROI) --- around the Q-value of the $0\nu\beta\beta$ decay transition, e.g., $\sim$3.0 MeV for $^{100}$Mo and $\sim$2.5 MeV for $^{130}$Te. The slant depth of the LSC Hall A, near the CROSS experimental place in the Hall B, under the highest summit M. Tobazo (1980~m) is 850~m and it can provide a valuable screening against muons \cite{Trzaska:2019}. 
However, the mountain profile under the LSC has a valley (Rioseta) with a minimal depth of $\sim$200~m to the LSC ground level, which reduces the efficiency of suppressing muons in the rock and results in a residual muon flux of 5.3~$\times$~10$^{-3}$~m$^{-2}$s$^{-1}$ with a dominant direction around the zenith angle of 40$^{\circ}$ and the azimuth angle of 150$^{\circ}$ \cite{Trzaska:2019}. Due to such a relatively high rate of muons in the LSC ($\sim$20~$\mu$/h/m$^2$), the muon-induced contribution to the ROI for $^{100}$Mo $0\nu\beta\beta$ decay search can be dominant, limiting the sensitivity of the CROSS experiment. Thus, a muon veto has been designed and installed around the CROSS cryogenic facility. In this paper, we describe the design, construction and operation of the muon veto system developed for the CROSS experiment (section \ref{sec:Veto_construction}), Monte Carlo simulations of muons in the CROSS setup to understand (section \ref{sec:Veto_Simulation}) and optimize (section \ref{sec:Veto_Optimization}) the muon veto, and a test of the veto in coincidence with two LMO cryogenic calorimeters (best-performance modules operated in the CUPID-Mo $0\nu\beta\beta$ decay search) used to validate the setup prior to the CROSS experiment (section \ref{sec:Veto_Validation}).

\section{Construction of the CROSS muon veto}
\label{sec:Veto_construction} 

\subsection{Design and construction}

\begin{figure}
\centering
\includegraphics[width=1.00\textwidth]{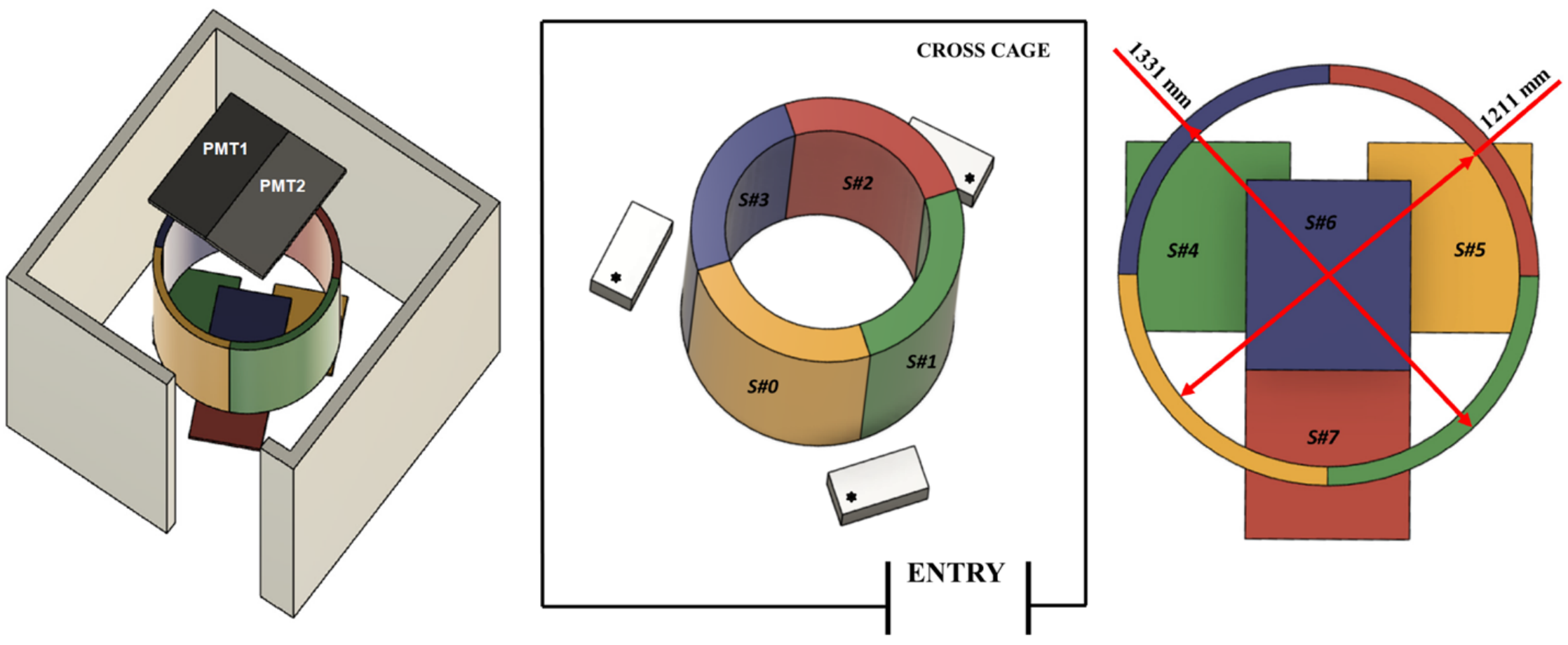}
\caption{Design of the main parts of the CROSS muon veto: 
(Left) View of the \emph{Top}, \emph{Lateral}, and \emph{Bottom} parts of the CROSS muon veto and the hut (without a roof for visual purpose), where the cryogenic facility is placed (not shown); 
(Middle) View on the the \emph{Lateral} veto with respect to the position of the three pillars of the cryostat and the hut's entrance;  
(Right) Top view on the \emph{Lateral} and \emph{Bottom} veto sectors.}
\label{fig:Muon_veto}
\end{figure}

The muon veto for the CROSS experiment, developed in the LNGS, consists of three main parts referred to as \emph{Lateral}, \emph{Bottom}, and \emph{Top}, which are schematically presented in figure \ref{fig:Muon_veto}. The \emph{Lateral} and \emph{Bottom} parts are installed inside the hut with the CROSS cryogenic facility, while the \emph{Top} part is placed on the roof of the hut.  

\begin{figure}
\centering
\includegraphics[width=0.35\textwidth]{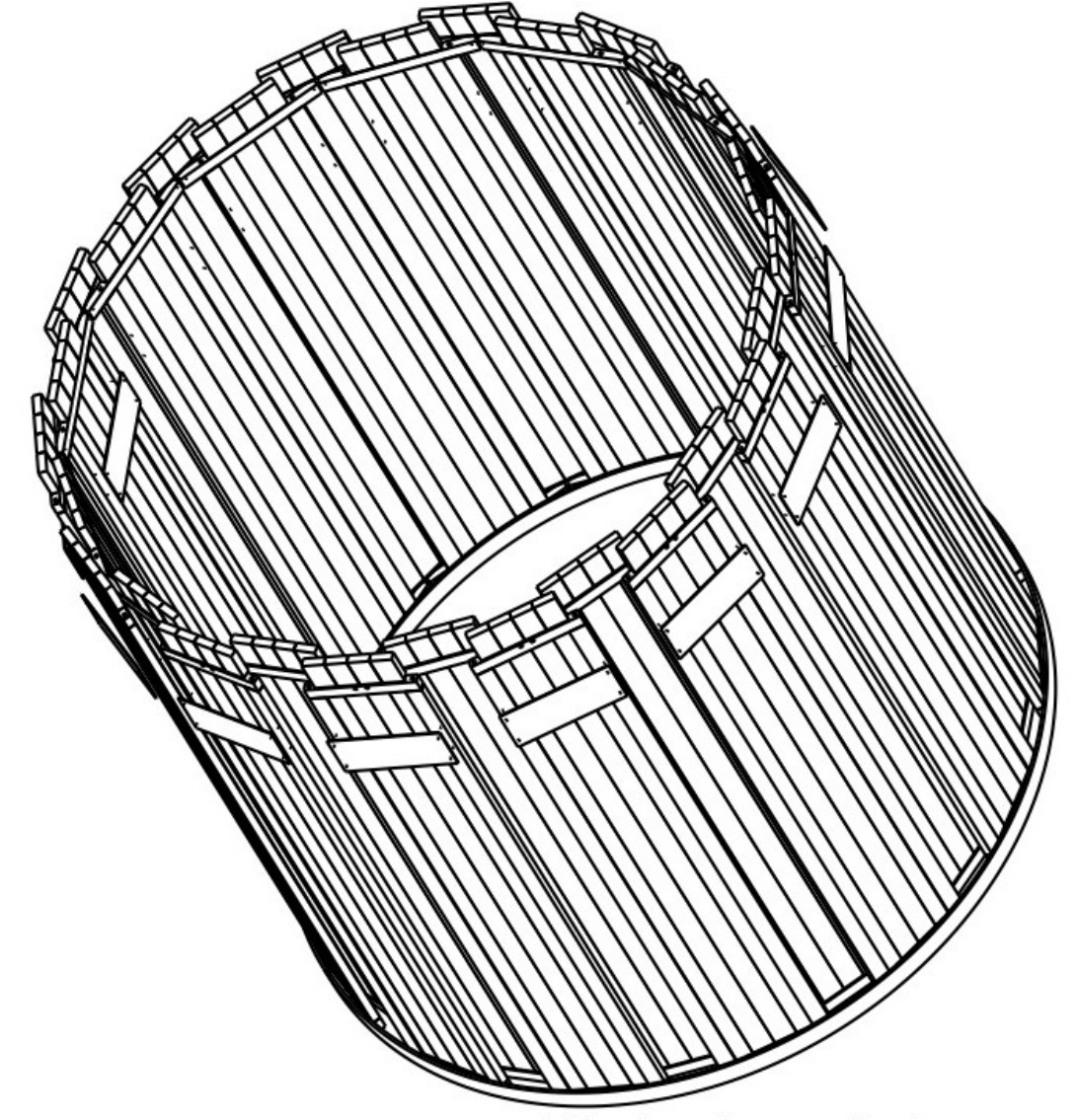}
\caption{Drawing of the \emph{Lateral} muon veto made of 28~scintillating detector modules, the readout of which is split to 4~sectors (S0, S1, S2, and S3 in figure \ref{fig:Muon_veto}) for the muon trigger logic; see details in text.}
\label{fig:Veto_Lateral_drawing}
\end{figure}

\begin{figure}
\centering
\includegraphics[width=0.90\textwidth]{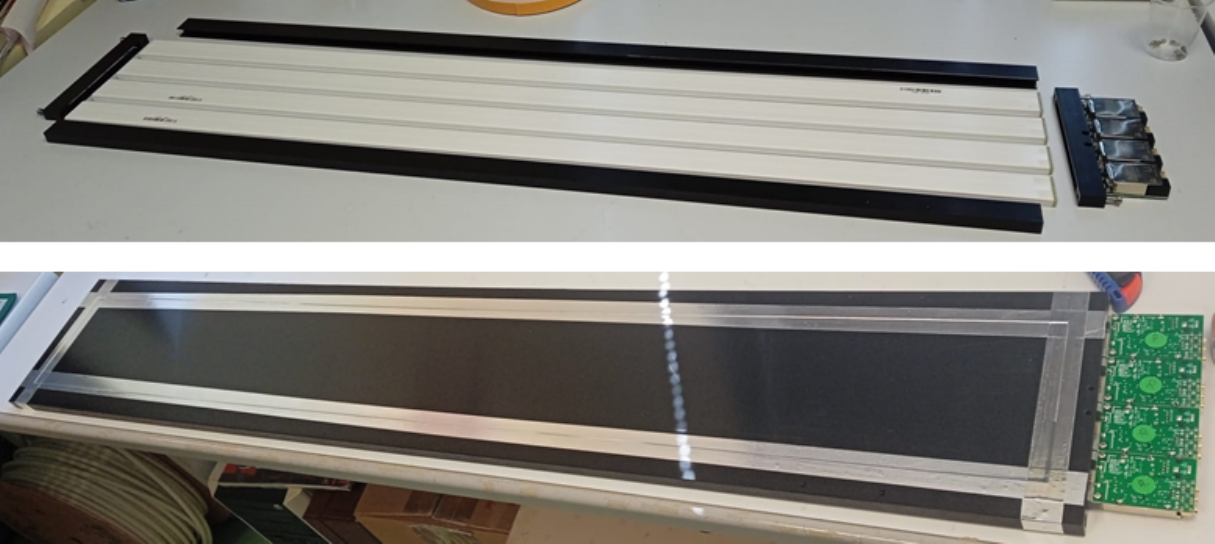}
\caption{Single section of the \emph{Lateral} veto before (top) and after (bottom) the assembly.}
\label{fig:Veto_Lateral_section}
\end{figure}

The \emph{Lateral} part consists of four sectors (called S0, S1, S2, and S3) that form a circle with an inner diameter of 121~cm (figure \ref{fig:Muon_veto}, right). According to the design of the \emph{Lateral} veto, shown in figure \ref{fig:Veto_Lateral_drawing}, each sector overlaps on the adjacent sectors by 1~cm. A single sector of the \emph{Lateral} veto consists of 7~modules made of 4~polystyrene bars (120~$\times$~4~$\times$~1~cm; provided by SCIONIX), each one optically coupled to a Si photomultiplier (SiPM; AdvanSiD, now Cefla, model ASD RGB 3S P), giving a total of 28~channels. 
A single module of the \emph{Lateral} veto system is shown in figure \ref{fig:Veto_Lateral_section}. 

\begin{figure}
\centering
\includegraphics[width=0.90\textwidth]{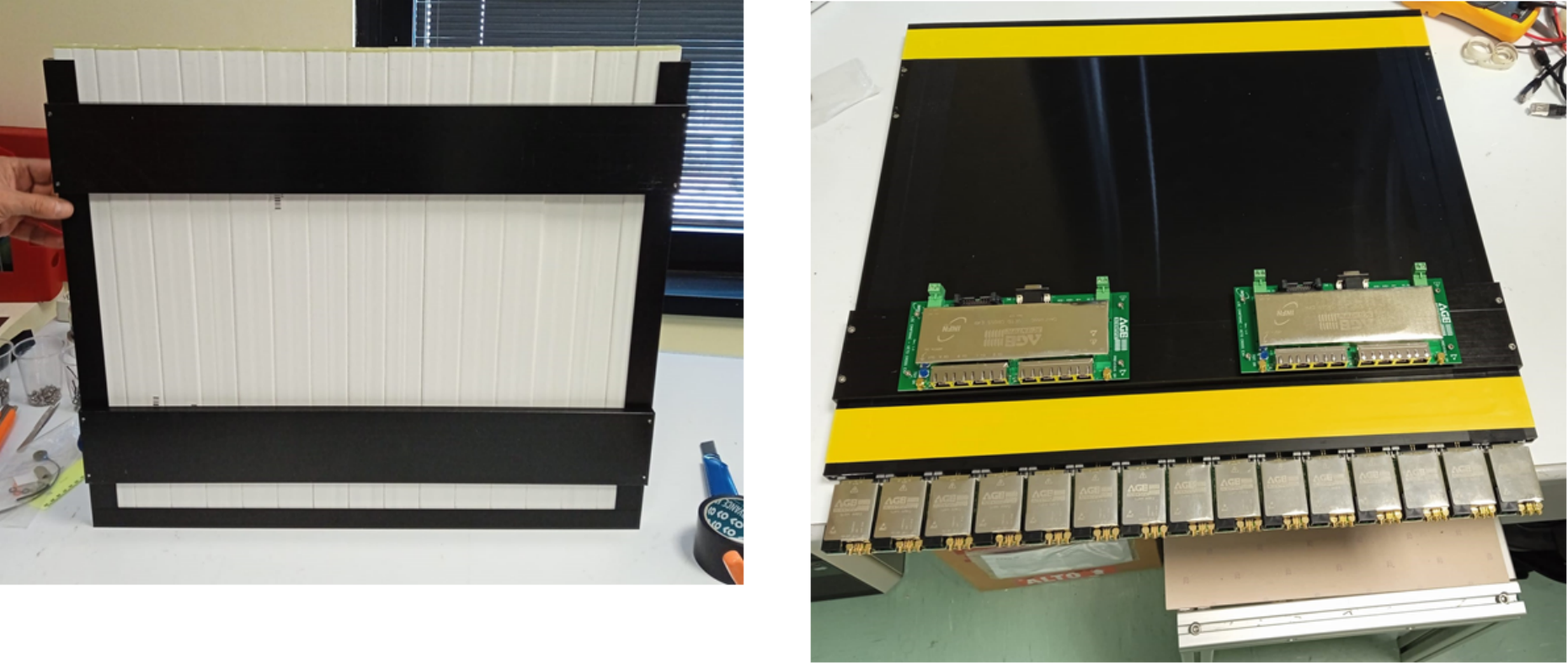}
\caption{Single panel of the \emph{Bottom} veto before (left) and after (right) the assembly. Splitter boards are the two long boards with eight RJ45 connectors, while the PCBs connected to the SiPMs are the 15~narrow boards on the bottom of the bars.}
\label{fig:Veto_Bottom_section}
\end{figure}  

\begin{figure}
\centering
\includegraphics[width=0.55\textwidth]{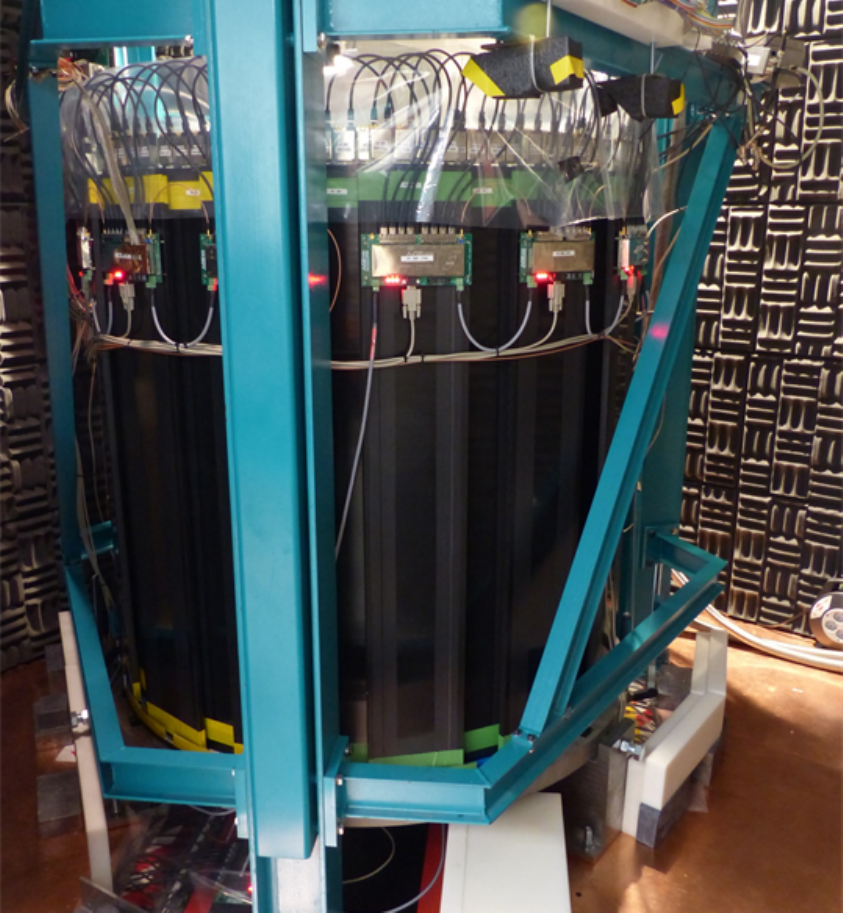}
\caption{Image of the CROSS setup, taken from the entry of the hut, showing a frame of the cryostat together with fragments of the \emph{Lateral} and \emph{Bottom} parts of the muon veto.}
\label{fig:Muon_veto_photo}
\end{figure}

The \emph{Bottom} part is also made up of four sectors (S4, S5, S6, and S7). Each sector consists of 15 polystyrene bars (52 $\times$~4~$\times$~1~cm), each one with a SiPM readout. 
Sector S6 is installed around 8~cm atop of other sectors, thus having a partial overlapping with them. A single sector of the \emph{Bottom} veto is presented in figure \ref{fig:Veto_Bottom_section}. The \emph{Lateral} and \emph{Bottom} parts of the CROSS muon veto system installed around the cryogenic facility are shown in figure \ref{fig:Muon_veto_photo}.

\begin{figure}
\centering
\includegraphics[width=0.53\textwidth]{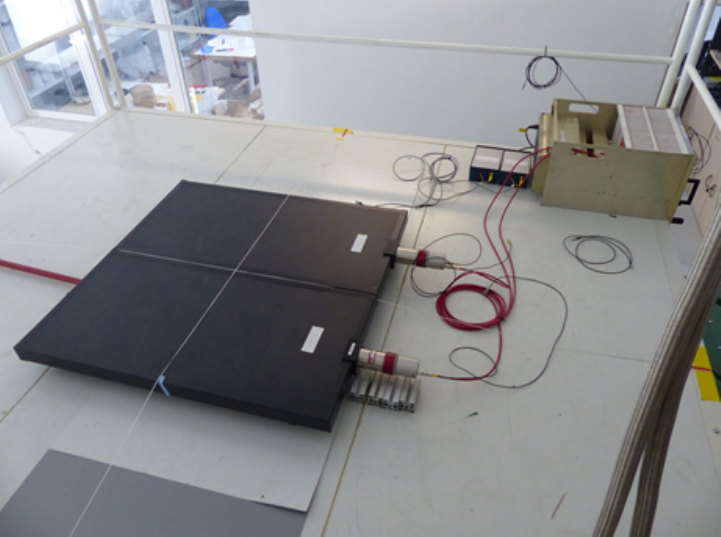}
\caption{View on the \emph{Top} part of the CROSS muon veto system installed on the roof of the hut. In addition to two veto panels, a CAEN crate with electronics is seen.}
\label{fig:Veto_Top}
\end{figure}

The \emph{Top} part of the CROSS muon veto, shown in figure \ref{fig:Veto_Top}, is formed from two 100~$\times$~50~$\times$~1~cm panels (S9 and S10) each readout with a photomultiplier (PMT; SCIONIX, model VS-0636-100). 
A combined channel of these two panels, S8, requiring a trigger of S9 OR S10, is acquired. 

\subsection{Electronics and DAQ}

\emph{Top}-veto PMTs are powered by HV through a crate (CAEN N470) and low voltage power supply, built at the LNGS, since they have an integrated preamplifier (SCIONIX VD14-E2). The signals from PMTs are inverted in a Phillips Scientific linear Fan in/out model 740 and then are fed into a discriminator (CAEN NIM model 96) that applies a threshold and generates a 0.5~ms OR logic pulse upon triggering. The OR signal and the PMT1 and PMT2 logic signals are stretched through 3 dual timers from CAEN and sent to VETO DAQ, in Muon VETO adapter, that receives trigger signals from PMT1, PMT2, PMT1 OR PMT2, and sectors S0--S7 of the \emph{Lateral} and \emph{Bottom} modules. 

\begin{figure}
\centering
\includegraphics[width=\textwidth]{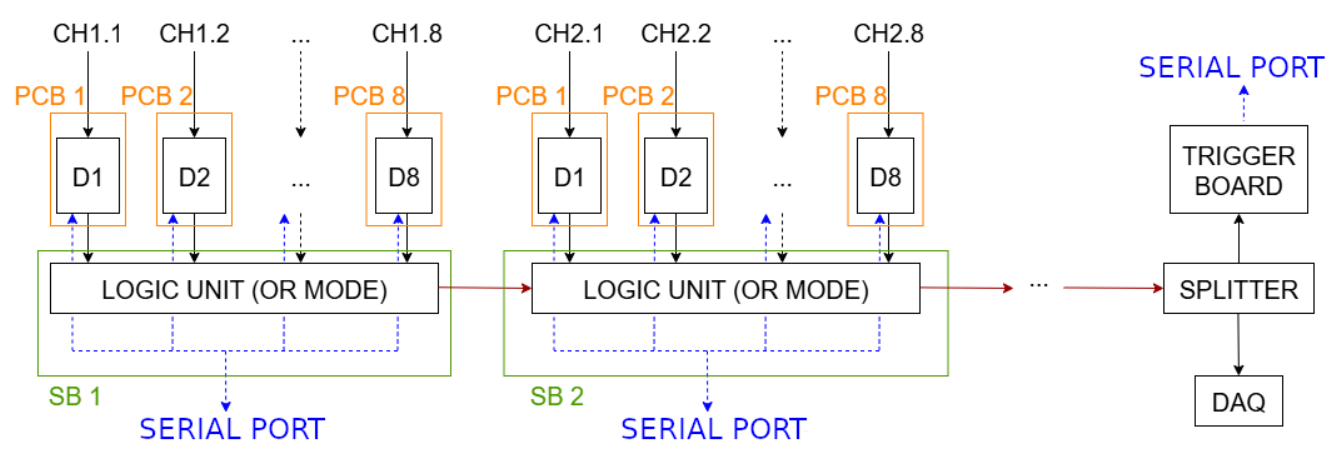}
\caption{Block diagram of the signal processing chain of the \emph{Lateral} and \emph{Bottom} sectors (see text). The discriminators (D1--D8) are contained in the SiPM AMP2, while the logic unit is included in the SB. The blue dashed lines represent the controller connection, which allows to set the bias of the SiPMs and the threshold level of the discriminator and to read the output of a single discriminator. The logic unit (working in OR mode of the eight discriminators plus the logic signal input) provides the output of each SB, shown in red color. The SBs of the same sector are connected in a daisy-chain. 
}
\label{fig:SBDAQ}
\end{figure}

The complete signal processing chain is shown schematically in figure~\ref{fig:SBDAQ}. Each SiPM of the \emph{Lateral} and \emph{Bottom} veto modules is connected to a dedicated PCB, called SiPM AMP2, which preamplifies the signal. It is connected to an 8-channel splitter board (SB) via Ethernet cable. Both the SiPM AMP2 and SB can be seen in figure~\ref{fig:Veto_Bottom_section}, right. The SB provides the SiPM bias, and generates a logic pulse upon triggering using a discriminator. The SBs are connected to a custom power supply that provides a separated channel for each sector. Each channel can be turned ON/OFF with a dedicated switch.
The SB includes a logic unit operated in OR mode, with nine inputs from the eight SIPM AMP2 plus an external logic input channel. 
This logic unit provides the output signal of the SB, which is a 0.5~ms-long square pulse. All SBs of the same sector are connected in a daisy-chain (the output of a SB is connected to the logic input of the next one). The output signal is split into two: one signal is sent to the DAQ and the other one to the trigger board.

Each SB has a serial port (24 in total) connected to a serial switch. 
A Raspberry Pi controller is plugged to that serial switch, and it can select via a Python script which serial port address will be configured. This controller allows to set the bias voltage applied to each individual SiPM as well as the threshold level applied to each signal. The bias of the SiPM can be configured from 19.2~V to 36.4~V in 256~steps of 67.5~mV each, while the threshold level can be set from 0~mV to 2.4~V in 64~steps of 37.5~mV each. Finally, another serial port connects the trigger board to the serial switch to acquire trigger patterns with an acquisition using the Raspberry Pi controller. 

The trigger board receives the output of the nine sectors and applies a logic trigger. The logic implemented on the board can be configured with a dedicated software, which allowed us to work with two different configurations: (1) a logical OR trigger of all sectors or (2) a trigger that requires a coincidence between two sectors in a 2~ms time window. 
Signals are converted in TTL differential signals and sent to DAQ through a receiver (muon veto receiver) plugged directly in the backplane. Another muon veto receiver, plugged in the second backplane connector, receives differential signals from triggers of single sectors. The receivers convert differential signal into single ended signals (S0--S7, PMT1, PMT2, PMT1 OR PMT2, and trigger) for DAQ. 

To reject muon-related events from the thermal detector data, it is essential to correlate them in time with events in the muon veto. For this purpose, two 8-bit FPGA GPI/O ports were dedicated to the muon veto in the CROSS DAQ, and acquired synchronously with data of thermal detectors (typically at 2 ksps). One of them registers the trigger pattern of the \emph{Lateral} and \emph{Bottom} sectors (S0--S7 obtained from the output of the last SB of a sector). This pattern is updated when the trigger board registers a new trigger. Only 4 bits are used in the second port: the outputs of the discriminator of each \emph{Top} module (PMT1 and PMT2), the output of the \emph{Top} logic unit (PMT1 OR PMT2), and the output of the trigger board. In the next, this last bit will be called the ``trigger bit''.

The rate of each individual channel of the \emph{Lateral} and \emph{Bottom} sectors (measured in one second) is also sent from the SBs to the Raspberry Pi controller through the serial ports. This allowed us to build a slow control of the system to ensure stability in the channels' rates and correct them in case of instabilities. A software in the controller measures the rate of each channel over one minute, and therefore it takes $\sim$3~h to check all of them. 
Once the rate measurement is completed, the rate of each channel is compared with the value of the rate set for that channel. If they differ more than 3$\sigma$, the threshold applied in the discriminator is changed $\pm$37.5~mV and the rate is measured again. This process is repeated until the difference is less than 3$\sigma$.

\section{Simulations of muon-induced background in the CROSS setup}
\label{sec:Veto_Simulation}

The Geant4~v11.1 toolkit~\cite{Agostinelli:2003} with the Livermore physics list was used to perform Monte Carlo (MC) simulations of muons flying through the CROSS setup. This section briefly describes the geometry implemented for the simulations, details the initial properties set to the generated muons, and outlines the obtained results. More details on the modeling and simulations of the CROSS setup together with low-temperature detectors can be found in a paper that describes the background projection for the CROSS experiment \cite{CROSS_Sensitivity:2025}.

\subsection{Geometry}
\label{sec:Geometry}

\begin{figure}
\centering
\includegraphics[width=1.00\textwidth]{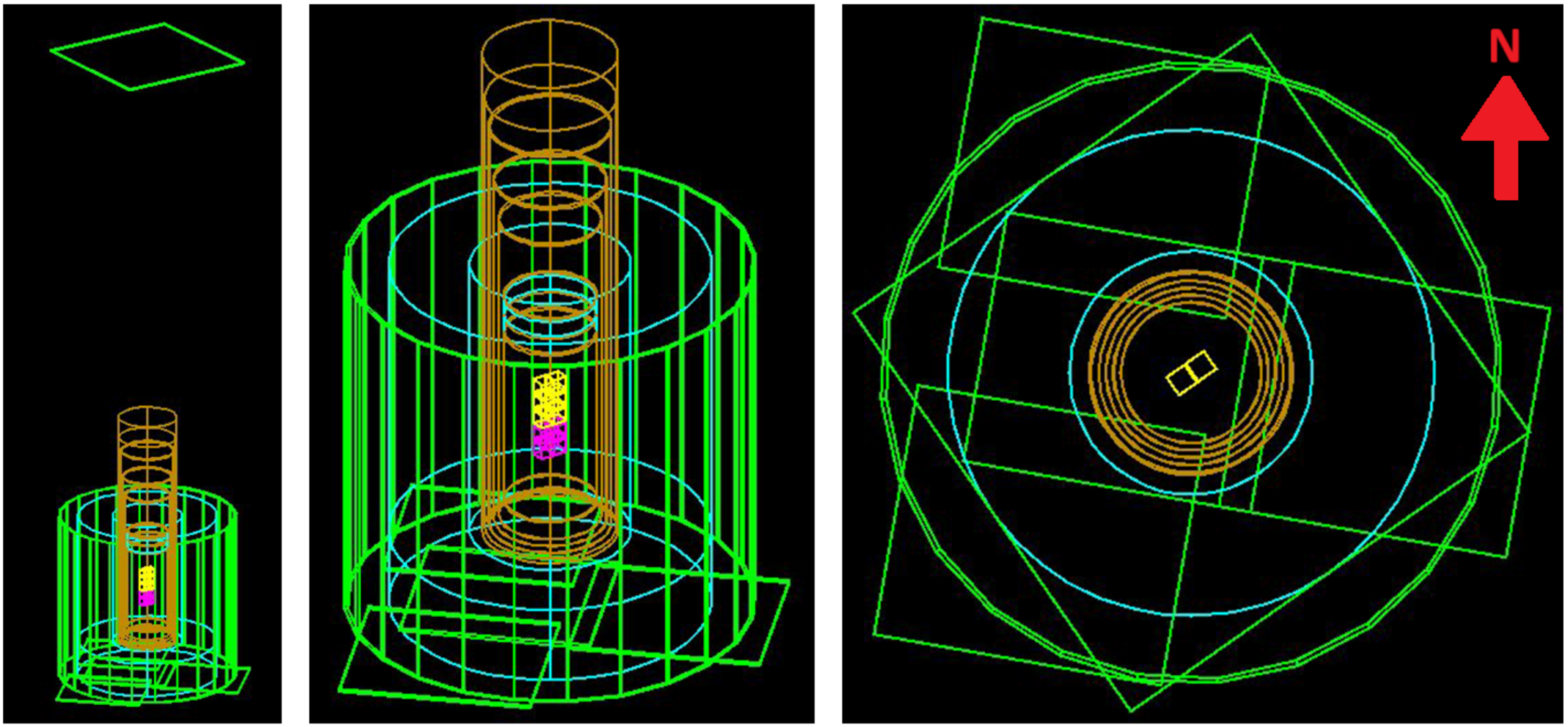}
\caption{Schematic view of the geometry implemented in the Geant4 simulation for the RUN~9 in the CROSS setup at the LSC (Spain). Green lines represent the muon veto, blue lines are the lead shielding, orange lines are the Cu thermal screens of the cryostat, yellow and purple lines are LMO and TeO crystals, respectively.}
\label{fig:MC_geometry}
\end{figure}

Figure~\ref{fig:MC_geometry} shows a schematic view of the geometry of the Geant4 simulation. The volumes simulated are six Cu thermal screens of the cryostat (simulated as empty cylinders), both internal and external lead shielding, the muon veto, and low-temperature detectors.

The muon veto has been simulated with an approximate shape as the real one without housing. To simplify the simulated geometry, it has been segmented into nine sectors. \emph{Lateral} sectors have been defined as one fourth of a 1-cm-thick cylinder, and no overlap has been considered in their geometry. Since the angular distribution of the muons is expected to be asymmetric \cite{Trzaska:2019}, they have been simulated with the real orientation they have in the laboratory. 

As explained above, the trigger for a sector can be done by any of the modules of this sector. Therefore, the energy deposit must be calculated in each module and not integrated in the sector. This granularity was implemented by integrating the energy deposits in each of the 174 modules considering their positions in the sector. Since the \emph{Bottom} and \emph{Top} sectors have the same shape as in the experiment, the positions of their modules are identical to the real ones. For the \emph{Lateral} sectors, each sector is divided into 28 identical modules.

To define a trigger in a given sector, the total energy deposited in one of their modules must be greater than a threshold level. If we require the coincidence between at least two sectors to make a trigger, then the time coincidence applied is 2~ms, the same as in the experiment (see details in section \ref{sec:RUN13_test}).

\begin{figure}
\centering
\includegraphics[width=0.3\textwidth]{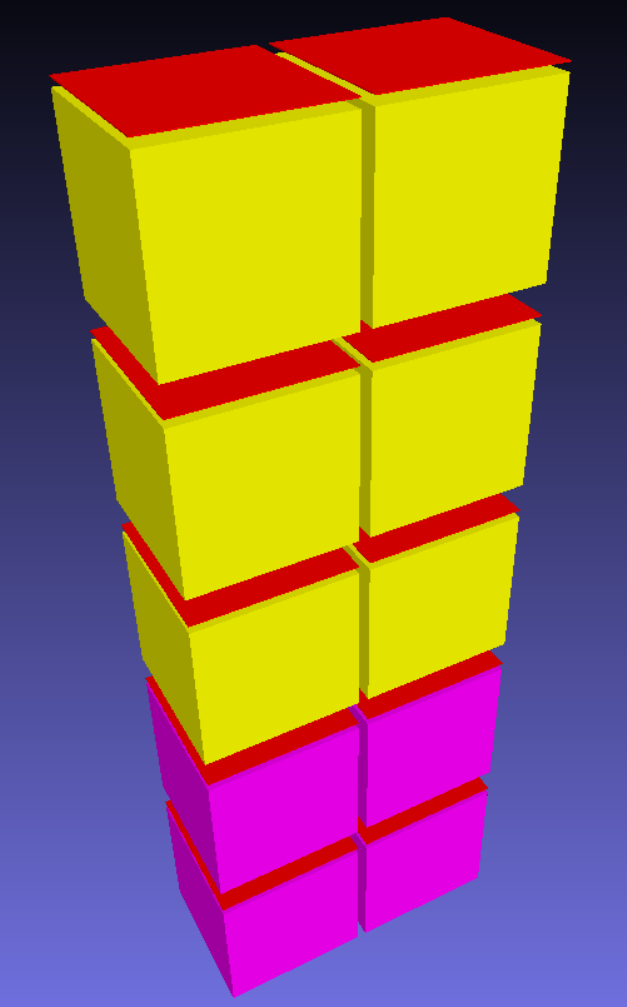}
\includegraphics[width=0.305\textwidth]{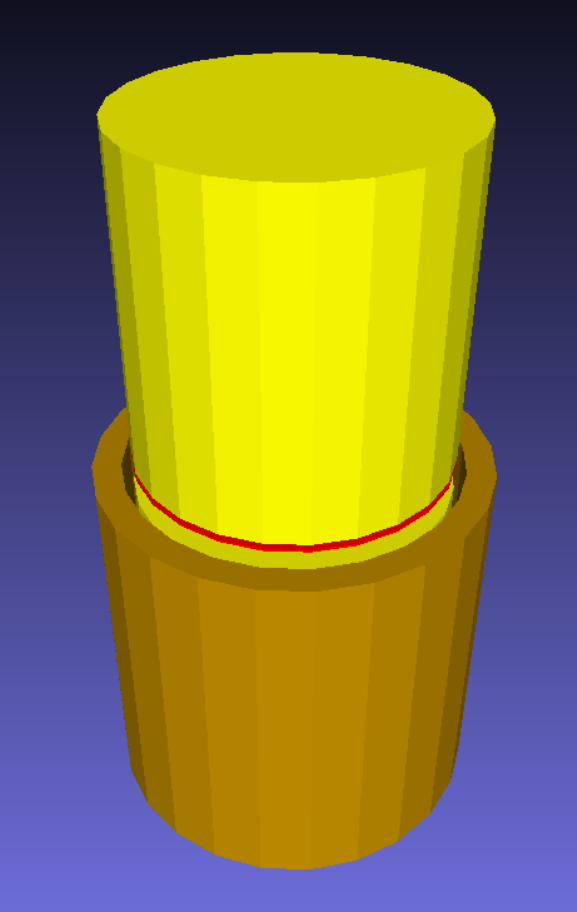}
\includegraphics[width=0.3\textwidth]{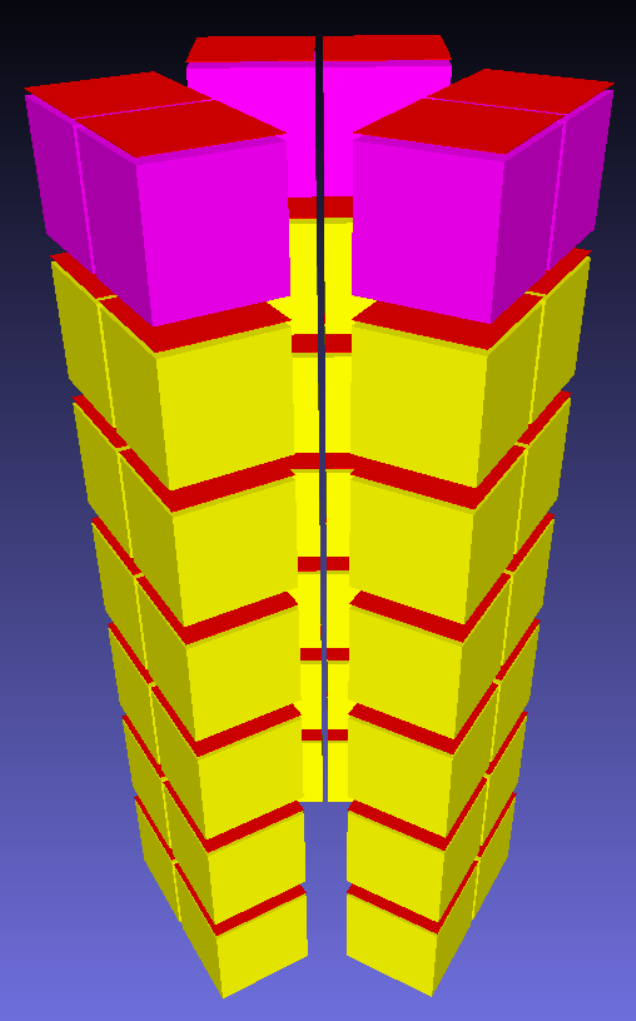}
\caption{Rendering of three Geant4 geometries of low-temperature detectors simulated inside the CROSS setup (see details in text): 
(Left) cryogenic RUN~9: 10-crystal tower composed by six LMOs and four TeOs, shown in yellow and purple color, respectively, and 10 Ge LDs in red;
(Middle) cryogenic RUN~13: a tower of two CUPID-Mo modules, composed by cylindrical LMOs and Ge LDs inside a copper housing, shown in yellow, red and orange, respectively. The Cu holder of the top module is not shown. 
(Right) CROSS experiment: three 14-crystal towers containing in total six TeOs placed on top of 36~LMOs; each crystal is accompanied with a cryogenic LD made of either Ge or Si.
}
\label{fig:MC_geometry2}
\end{figure}  

Three configurations of low-temperature detectors have been simulated (see in figure~\ref{fig:MC_geometry2}). 
The first simulated detector structure, used in measurement RUN~9 \cite{CROSS_enriched_TeO:2024,CROSS_Run9:2025}, is a 10-crystal tower arranged on five floors of two crystals each. It was composed by six LMOs (two of them are $^{100}$Mo-enriched) on top of four $^{130}$Te-enriched TeOs, as shown in figure~\ref{fig:MC_geometry2}, left. All the crystals are cubic with a 45-mm-side, and each one was accompanied with a light detector (LD) made of a 0.3-mm-thick 45-mm-side Ge wafer. The measurement campaign carried out with this tower and the results obtained are detailed in \cite{CROSS_enriched_TeO:2024,CROSS_Run9:2025}. In the same way as for the muon veto, these detectors have been simulated with the real orientation they had in the laboratory during the measurements. This orientation can be clearly seen in figure~\ref{fig:MC_geometry} (right). 
The second detector configuration defined corresponds to the measurement RUN~13, the last cryogenic run of the setup prior to the CROSS experiment, in which a mini-tower of two CUPID-Mo modules originally used in the CUPID-Mo experiment~\cite{Armengaud:2020a} was operated (see Sect. \ref{sec:RUN13_test}). Each module contains a cylindrical LMO crystal ($\oslash$44 $\times$ 45~mm) and a LD made of a 175-$\mu$m-thick Ge wafer with the same diameter as the LMO. Both detectors were placed inside a 2-mm-thick copper housing. 
Finally, an approximate geometry of the CROSS detector array \cite{CROSS_Sensitivity:2025} was simulated. It consists of three 14-crystal towers distributed in seven floors of two crystals each, containing six $^{130}$Te-enriched TeO crystals placed on top of the 36 LMOs (32 of them are $^{100}$Mo-enriched). Each crystal is accompanied with a square-shaped (45-mm side) LD made of Ge or Si 0.3-mm-thick wafers. 
In all configurations, an LD placed between two crystals can detect scintillation signals of both.

\subsection{Muon angular distribution}
\label{sec:MuonDistribution}

\begin{figure}
\centering
\includegraphics[width=0.90\textwidth]{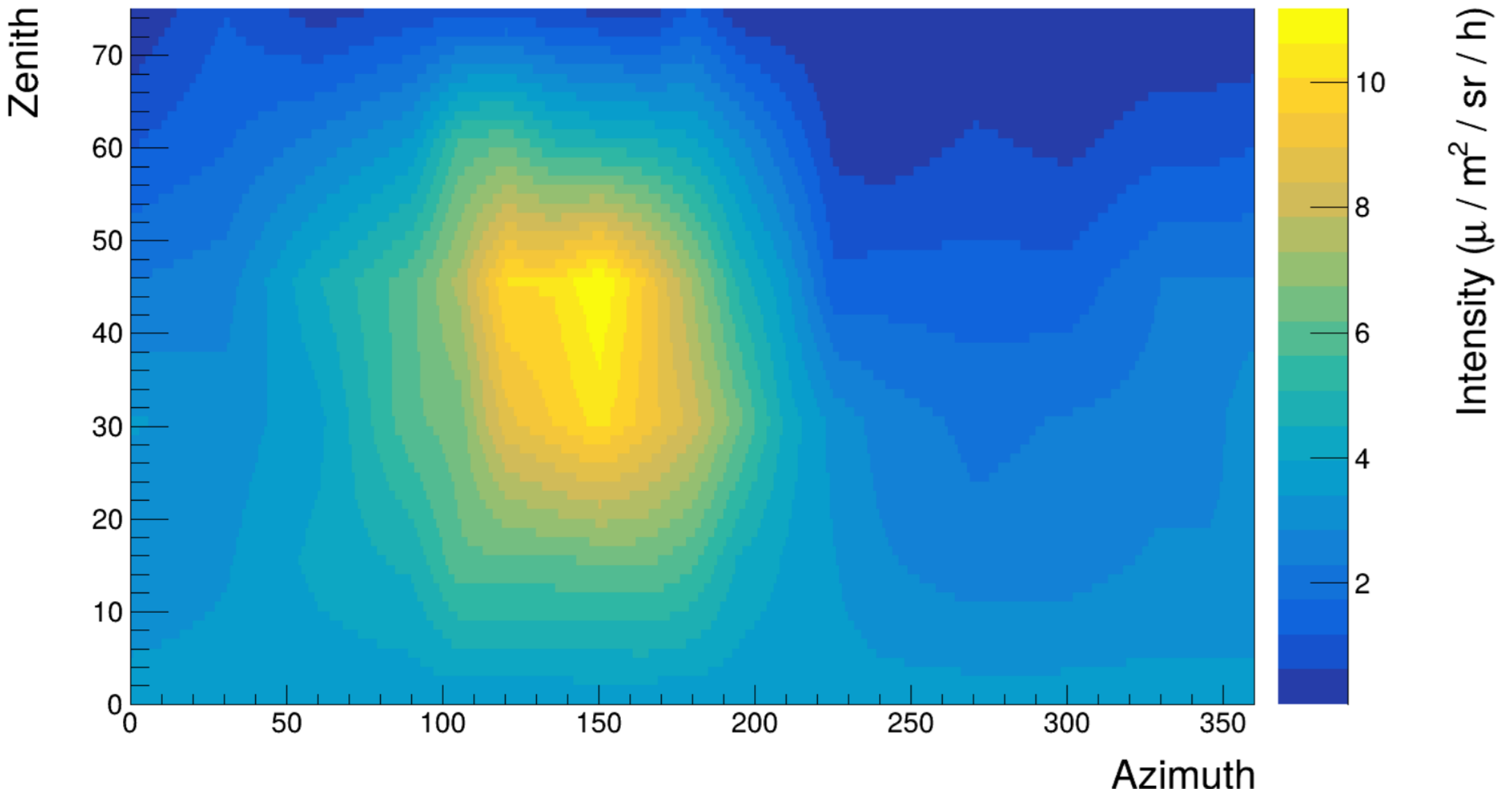}
\caption{Parametrization of muon's angular distribution at the LSC Hall~A implemented in the Geant4 MC simulations of muon-induced events in the CROSS setup.}
\label{fig:Muon_parametrization}
\end{figure}

The flux and angular distribution of muons reaching the Hall~A of the LSC were measured with a Muon Monitor \cite{Trzaska:2019}, reporting the muon flux of around 19(1)~$\mu$/m$^2$/h and its predominant direction at an azimuth angle of $\phi$ = 150$^\circ$ and a zenith angle of $\theta$ = 40$^\circ$. This direction points to the Rioseta valley. Thus, in order to account properly for the muon flux in the CROSS setup, we parameterize muons according to the results of Ref. \cite{Trzaska:2019} and the two-dimensional ($\phi$ $\times$ $\theta$) muon angle distribution obtained is shown in figure~\ref{fig:Muon_parametrization}. It should be emphasized that the CROSS facility is installed in the Hall~B, which is $\sim$20~m closer to the Rioseta valley than the Hall~A, and therefore it is expected to be irradiated with slightly higher muon flux. However, due to the lack of measurements in Hall~B, we used those presented in~\cite{Trzaska:2019} to get the initial angles of the simulated muons, which were randomly selected from the parameterized two-dimensional angular distribution. Since $\theta$ of muons was measured up to 75$^\circ$, this is also the maximum angle of muons simulated.

\subsection{Slant depth}
\label{sec:SlantDepth}

\begin{figure}
\centering
\includegraphics[width=0.90\textwidth]{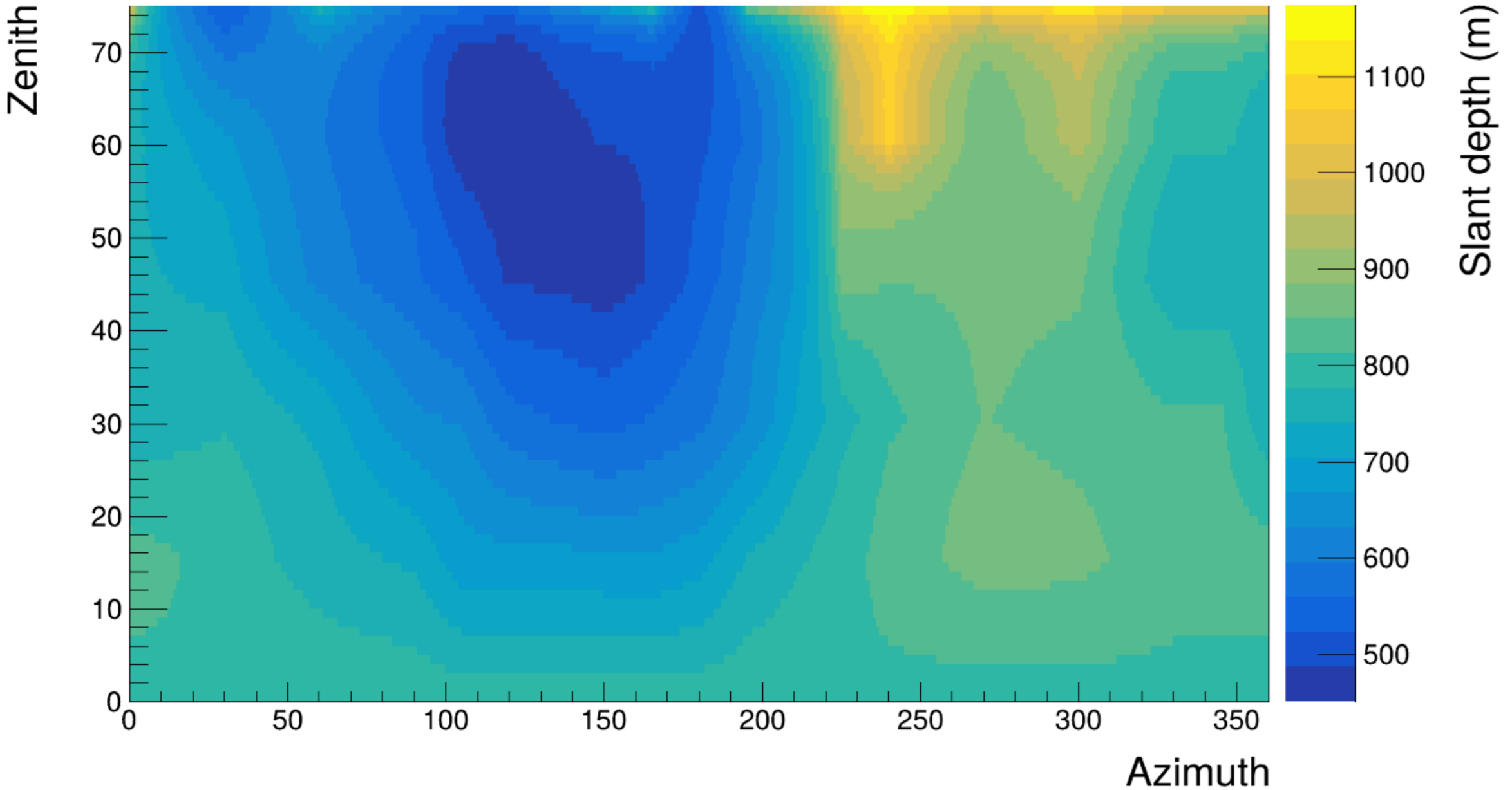}
\caption{Parametrization of the slant depth under the LSC Hall~B, where the CROSS setup is located.}
\label{fig:Depth_LSC}
\end{figure}

In~\cite{Trzaska:2019}, the muon flux measured in the LSC was fitted to a semi-empirical dependence on the slant depth $x$~\cite{Mei:2006}:
\begin{equation}
    I(x) \approx I_1 ~e^{-x/\lambda_1} + I_2 ~e^{-x/\lambda_2},
\end{equation}
with the following fit parameters 
$I_1$ = 8.60(53) $\times$ 10$^{-6}$ cm$^{-2}$s$^{-1}$sr$^{-1}$,
$I_2$ = 0.44(6) $\times$ 10$^{-6}$ cm$^{-2}$s$^{-1}$sr$^{-1}$, 
$\lambda_1$ = 450(10) m.w.e. (m of water equivalent) and $\lambda_2$ = 870(20) m.w.e. This approximation was used to obtain the rock depth as a function of the angle ($\phi$ and $\theta$) using the angular distribution measured in the LSC. Since the intensity of muons reaching the surface of the mountain follows the $\cos^2$($\theta$) law, this has also been taken into account as a reduction of the intensity at high zenith angles. The resulting parameterization is shown in figure~\ref{fig:Depth_LSC}.

\subsection{Initial position and momentum of muons}

\begin{figure}
\centering
\includegraphics[width=0.5\textwidth]{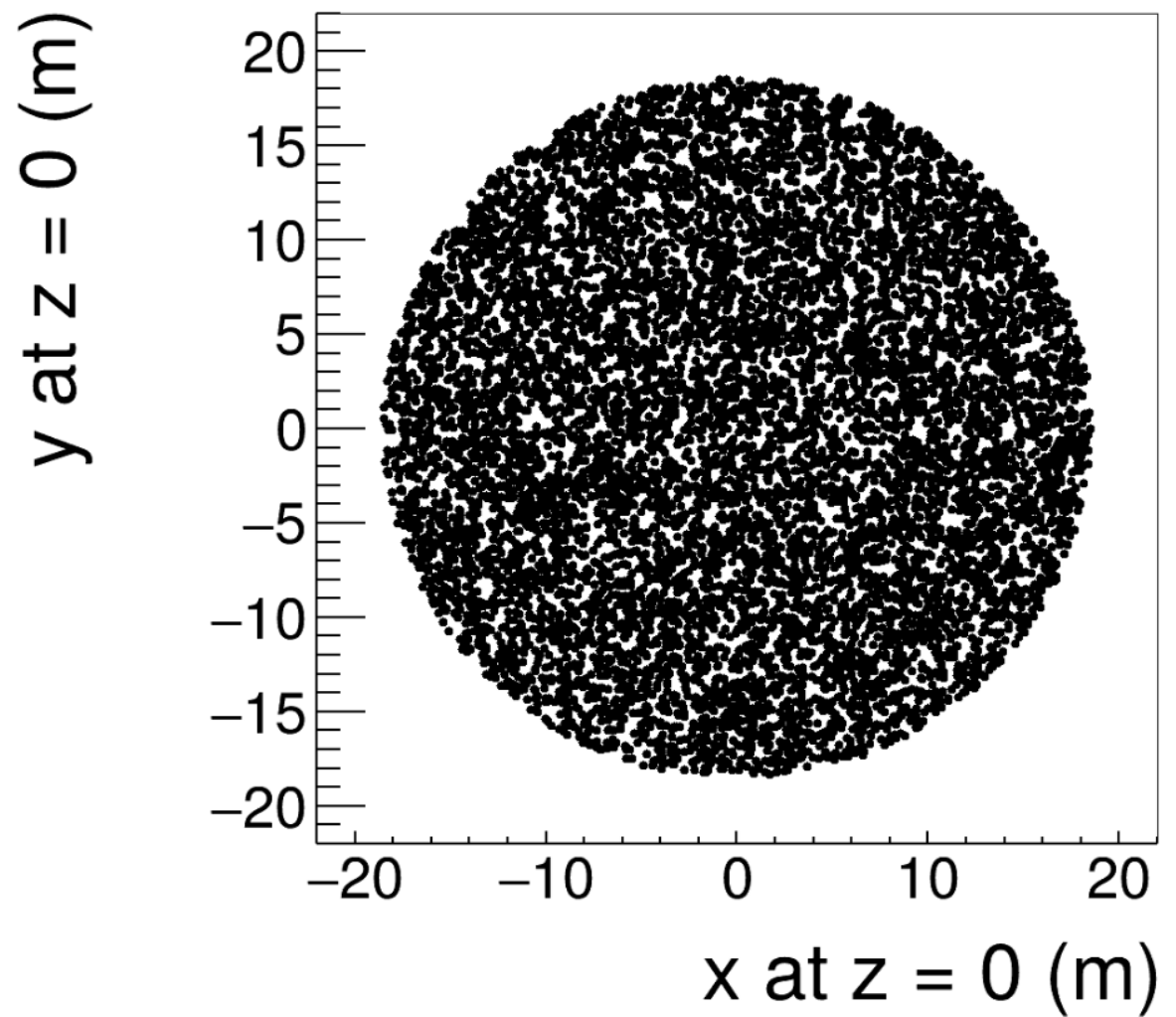}
\caption{X:Y position in the Z = 0 plane of 10$^4$ simulated muons.}
\label{fig:x0y0Muons}
\end{figure}

To be able to calculate the event rates of muons in the detectors using the muon flux, it is essential to ensure a homogeneous flux within the experimental setup. For this purpose, both the position and momentum of the muons have been chosen to allow them to cover a circular surface in the Z = 0 plane (just below the experiment) large enough to fully shine the experiment with muons up to the simulated maximum zenith angle (75$^\circ$). Therefore, the radius of that circle was:
\begin{equation}
    r_{\textrm{MC}} = r_{\textrm{Exp}} + h_{\textrm{Exp}} \cdot \tan(\theta_{\textrm{max}}),
\end{equation}
where $r_{\textrm{Exp}}$ is the radius of the experimental setup ($\sim$80~cm), $h_{\textrm{Exp}}$ is the total height ($\sim$4.8~m) and $\theta_{\textrm{max}}$ is the maximum zenith angle. To allow muons to fully irradiate the experimental setup, their initial positions were randomly selected within a circle in the Z = D plane, and D was chosen to be larger than $h_{\textrm{Exp}}$. The center of the circle is determined by the zenith and azimuth angles obtained in section~\ref{sec:MuonDistribution}: X$_0$ = D $\cdot\tan$($\theta$) $\cdot\sin$($\phi$), Y$_0$ = D $\cdot\tan$($\theta$) $\cdot\cos$($\phi$), Z$_0$ = D. The radius of the circle is also $r_{\textrm{MC}}$. The initial momentum of the muons is then determined by the zenith and azimuth angles in such a way that there is a unique correspondence between the initial position of the muon in the Z = D plane and the position that it would have in the Z = 0 plane if there was nothing to interact with. That is $p_{\textrm{X}_0}$ = $-\sin$($\theta$) $\cdot\sin$($\phi$), $p_{\textrm{Y}_0}$ = $-\sin$($\theta$) $\cdot\cos$($\phi$), $p_{\textrm{Z}_0}$ = $-\cos$($\theta$). Figure~\ref{fig:x0y0Muons} shows a homogeneous distribution of muons reaching the Z = 0 plane.

\subsection{Initial energy of muons}

For each simulated muon, its initial energy at sea level, $E_{\mu,0}$, has been randomly selected from a probability density function (PDF) defined in \cite{Workman:2022} as:
\begin{equation}
    \frac{dN_\mu}{dE_{\mu,0} ~d\Omega} =  \frac{0.14 ~E^{-2.7}_{\mu,0}}{\textrm{cm}^2 ~\textrm{s} ~\textrm{sr} ~\textrm{GeV}} \cdot \Biggl\{ \frac{1}{1+\frac{1.1 ~E_{\mu,0} ~\textrm{cos}\theta}{115 ~\textrm{GeV}}} + \frac{0.054}{1+\frac{1.1 ~E_{\mu,0} ~\textrm{cos}\theta}{850 ~\textrm{GeV}}} \Biggl\} .
    \label{eq:Muon_distribution}
\end{equation}
This is an approximation for muons with energies higher than 100~GeV on the surface level. Since all muons reaching the laboratory should have much larger energies aboveground, this approach is suitable for this simulation. Once the energy of a muon on the surface is selected, the energy of that muon underground, $E_{\mu}$, is calculated using a relation between the energy of the muon at the surface and its energy underground \cite{Workman:2022}:
\begin{equation}
    E_{\mu} = (E_{\mu,0} + \varepsilon) \cdot e^{-b \cdot x}-\varepsilon,
    \label{eq:Muon_energy}
\end{equation}
where $x$ is the slant depth obtained in section~\ref{sec:SlantDepth}, $\varepsilon\approx$~2.5~TeV, and $b^{-1}\approx$~2500~m.w.e are parameters of the rock overburden.

\subsection{Output of simulations}

The information provided by the simulation is the total energy deposited in each detector (veto modules, crystals, and LDs) by each kind of particle ($\mu$, $\beta$/$\gamma$ or $\alpha$), or by any particle and the triggering patterns and multiplicities above the energy threshold in each kind of detectors. The initial properties (energy, angles, and position) of each simulated muon are also recorded. Only events with energy deposit in the detectors have been saved. The muon flux measured in the LSC has been used to build the rates of each detector and the normalized energy spectra.

\subsubsection{Validation of simulations}
\label{sec:MC_validation}

To validate MC simulations, 3 $\times$ 10$^9$ muons were generated for the RUN~9 configuration (see section \ref{sec:Geometry}) \cite{CROSS_Run9:2025}, and the rates in the detectors were compared with the experimental data (112 h of measurements). In the experimental data, muon-like interactions have been selected in six LMO thermal detectors as those with an energy deposit >~10~MeV and the scintillation light expected for a $\beta$/$\gamma$ interaction, thus giving a rate of 2.5(4)~cnts/d/LMO. The simulation agrees with this result, since the rate of events with energy deposit in crystals >~10~MeV and no interaction with the LDs (i.e., they only absorb scintillation light from crystals) is 2.6(1)~cnts/d/LMO.

\begin{figure}
\centering
\includegraphics[width=0.8\textwidth]{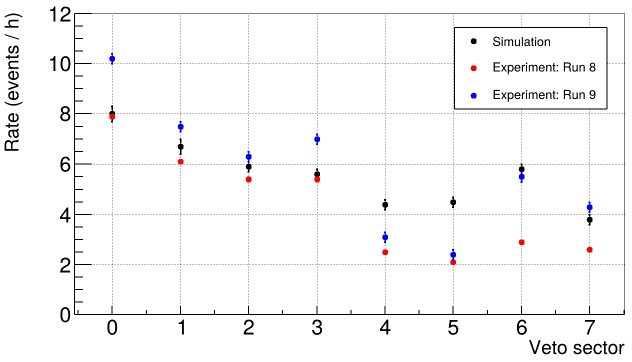}
\caption{Simulation and experimental comparison of the rates of the muon veto sectors for coincident events in at least two sectors. Two different experimental runs are compared: RUN~8 (red points) and RUN~9 (blue points), corresponding to time exposures of 911~h and 612~h, respectively. An energy threshold of 1~MeV in all the veto modules was considered in the simulation.}
\label{fig:MuonRateVetos}
\end{figure}

The rate in the \emph{Lateral} and \emph{Bottom} veto sectors predicted by simulations for coincident events in at least two sectors is compared to two measurements at the LSC, RUN~8 \cite{CROSSdetectorStructure:2024} and RUN~9 \cite{CROSS_enriched_TeO:2024,CROSS_Run9:2025} (with time exposures of 911~h and 612~h, respectively), as shown in figure~\ref{fig:MuonRateVetos}. In the simulations, an energy threshold of 1~MeV was assumed for each veto module. As seen in this figure, the experimental rates do not agree with each other, which can be explained if the energy thresholds of the veto modules are not homogeneous or if they change with time. This indicates that the setup of the veto must ensure a homogeneous energy threshold, achieved in the following studies described in section~\ref{sec:Veto_Validation}. On the other hand, the simulation agrees with the experimental data in the fact that the rate of the \emph{Lateral} sectors is higher than that of the \emph{Bottom} ones, due to the angular distribution of the muons and the surface covered by these sectors. This can be clearly seen in the rate of the sector S0, which is higher than the others, since it is directly pointing to the Rioseta valley.

\subsubsection{Muon-induced background in the CROSS setup}

\begin{figure}
\centering
\includegraphics[width=0.7\textwidth]{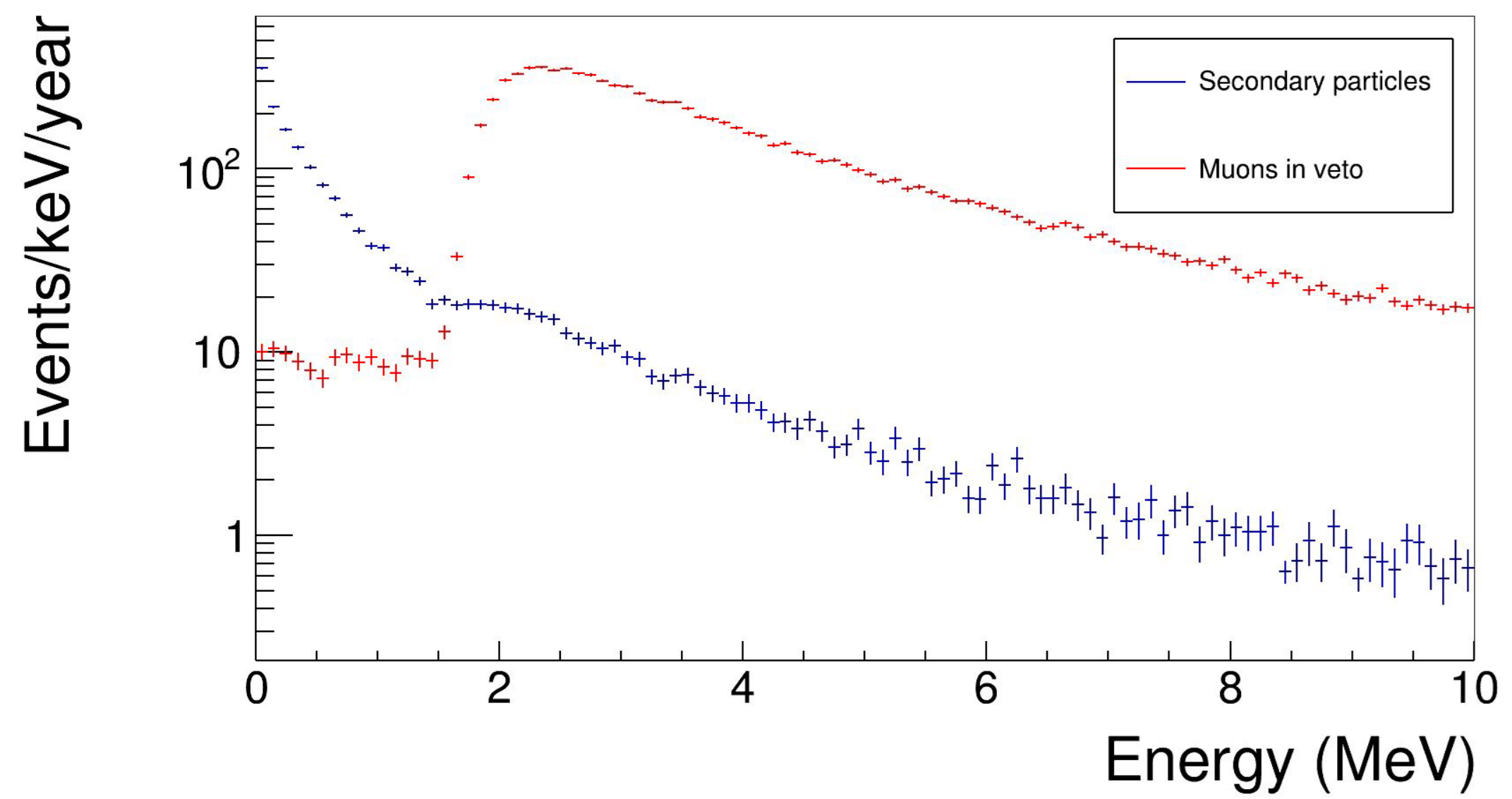}
\includegraphics[width=0.7\textwidth]{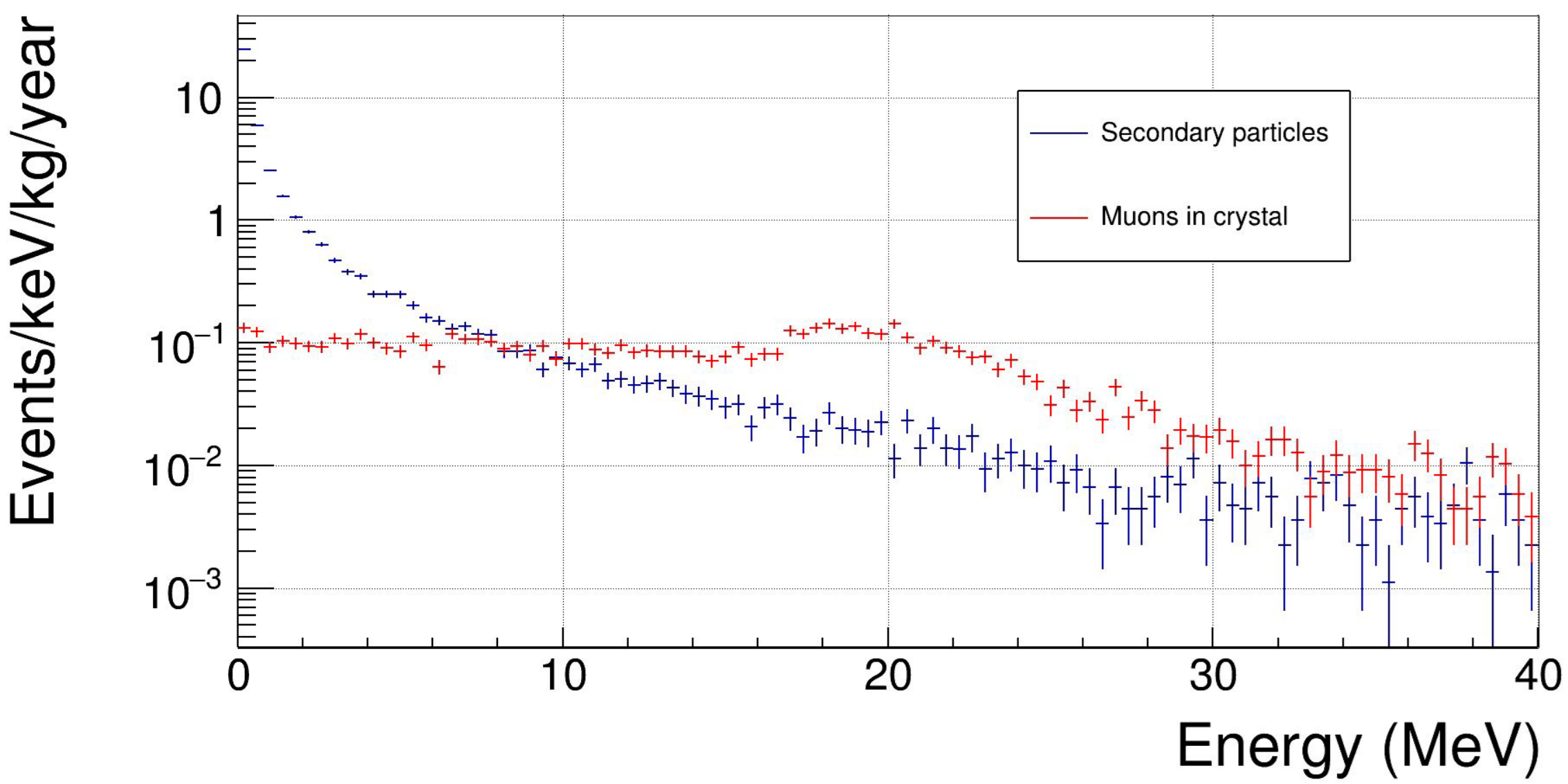}
\caption{Energy spectra of primary and secondary events induced by muons passing through the CROSS muon veto (top) and detected by the LMO thermal detector (bottom).}
\label{fig:Muon_induced_events}
\end{figure}

The output information of the simulations allows us also to compare the energy spectrum of events where the muon interacted with the detector with the spectrum of events where the energy was deposited by a secondary particle produced in another volume of the setup after a muon interaction. Figure~\ref{fig:Muon_induced_events} shows the comparison of these spectra for both a single muon veto sector and an LMO thermal detector.

The muon veto spectra presented in figure~\ref{fig:Muon_induced_events} (top) show that the muon bump starts at $\sim$1.5~MeV, and this fixes the maximum energy threshold for the veto modules that will allow us to detect most muons. It is worth noticing that since the main part of the muon bump is within the range of the natural radioactivity background ($<$2.6~MeV), thus a contribution of no muon-related events triggering the veto modules is expected. Since the trigger of a sector can be done by any of its modules, this contribution can be very significant compared to muon-related events.

\begin{figure}
\centering
\includegraphics[width=0.75\textwidth]{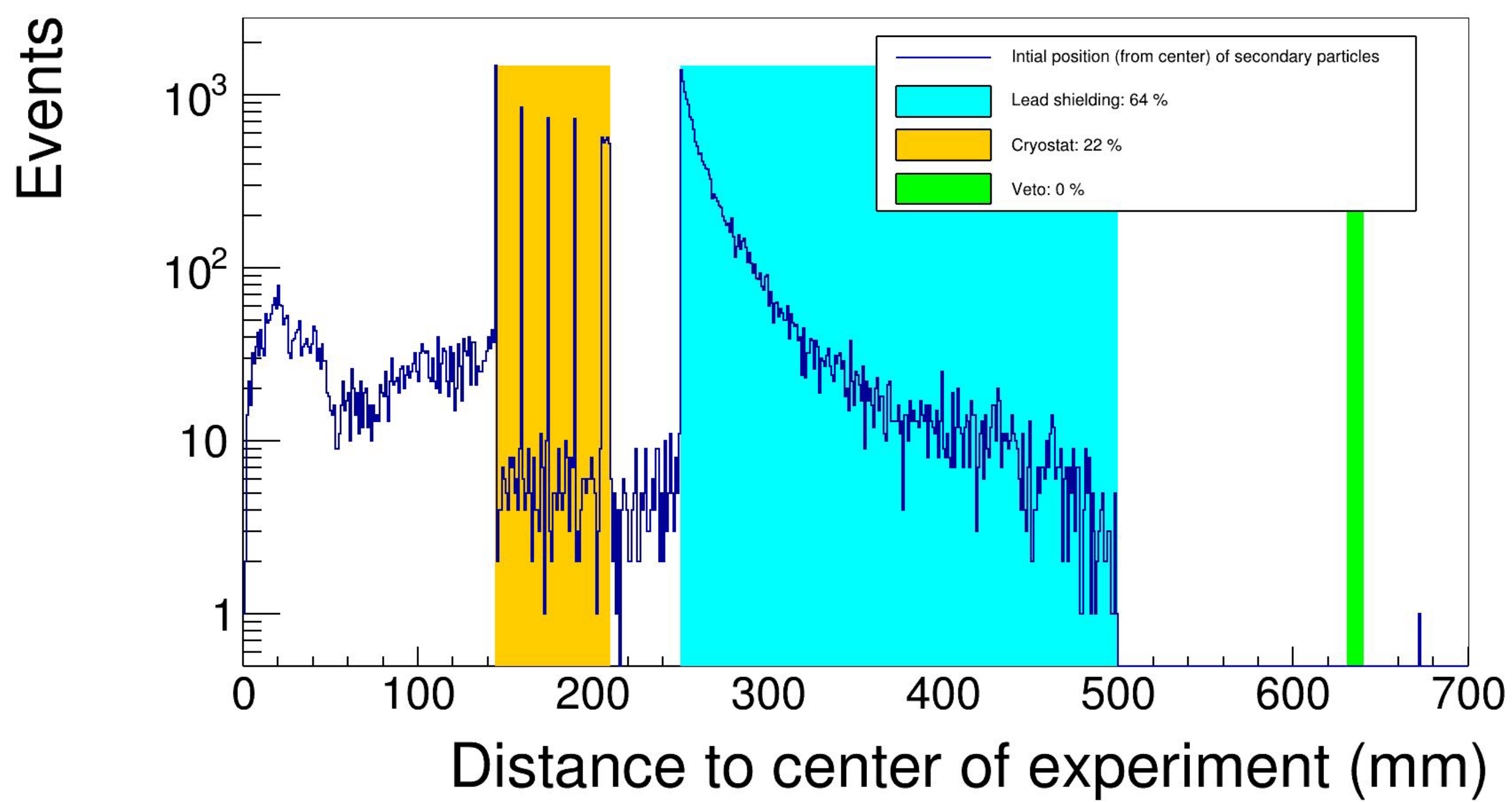}
\caption{Production of muon-induced secondary particles on the construction materials of the CROSS setup which subsequent energy deposition in an LMO thermal detector.}
\label{fig:Muon_induced_secondary_evts}
\end{figure}

In case of the LMO detector (figure \ref{fig:Muon_induced_events}, bottom) secondary particles are dominant at energies below $\sim$6~MeV, and without any event selection they contribute to a rate of $\sim$0.6~cnts/keV/kg/yr in the ROI. This background contribution is around three orders of magnitude higher than that required for the CROSS experiment to achieve the projected sensitivity \cite{Bandac:2020}. Therefore, it is essential to understand the origin of these events and try to reduce their contribution. The simulation showed that $\sim$90\% of the events were initiated by $\beta$ particles or $\gamma$ quanta produced in the volumes surrounding the detectors. Figure~\ref{fig:Muon_induced_secondary_evts} demonstrates the origin position of the secondary particles with respect to the center of the experiment. This allows us to identify that 64\% of these particles are produced in the lead shield, while other 22\% originate in the Cu screens of the cryostat and the rest is ascribed to the detector materials (mainly, the Cu structure). This means that it is not possible to shield detectors inside the CROSS setup from these particles, and the only way to reject them is to tag muons passing through the facility. 

To achieve a minimum acceptable background contribution of muon-related events in the crystals, different event selections have been taken into account to minimize the presence of any kind of background event in the analysis while maintaining a high acceptance of $0\nu\beta\beta$ events. First, the $0\nu\beta\beta$ decay is localized within just one crystal, and therefore the events must be single-crystal hit. Moreover, the energy deposited in the LDs must correspond to the scintillation light emitted by $\beta$/$\gamma$ particles. In the simulation, this implies to reject events with a multiplicity~>~1 in the crystals ($M_B>1$) and events with energy deposit in any LD ($M_{LD}>0$). Moreover, two different configurations have been considered for the muon veto: rejection of coincident events across multiple sectors ($M_V>1$) or rejection of events with energy deposit in any veto sector ($M_V>0$). The background index of muon-induced events expected in the CROSS experiment depending on the event rejection strategy is listed in table~\ref{tab:EventSelections}. An energy threshold in the muon veto of 1~MeV has been considered. To calculate these contributions, the total number of events that fulfill the event selections have been integrated between 2.7 and 3.4 MeV, since the ROI of the experiment is centered at 3.034~MeV.

\begin{table}
    \caption{Background contributions in the 2.7--3.4 MeV interval, including the $^{100}$Mo $0\nu\beta\beta$ ROI (3.034~MeV), obtained for the CROSS experiment configuration without and with the rejection of muon-induced coincidences. An energy threshold in the muon veto of 1~MeV is assumed.}
    \centering
    \begin{tabular}{c|c}
    \hline
        Event rejection & BI (10$^{-3}$ cnts/keV/kg/yr) \\ \hline
        No rejection & 653(13) \\
        $M_B>1$ OR $M_{LD}>0$ & 8.5(1) \\
        $M_B>1$ OR $M_{LD}>0$ OR $M_V>1$ & 5.0(1) \\
        $M_B>1$ OR $M_{LD}>0$ OR $M_V>0$ & 1.8(1) \\
        \hline
    \end{tabular}
    \label{tab:EventSelections}
\end{table}

Since muons produce a shower of secondary particles when they interact in the volumes surrounding thermal detectors, the most relevant event selection is a multiplicity cut. With this event selection, a reduction of 98.7\% of the events can be achieved in the CROSS experiment (see table~\ref{tab:EventSelections}). Despite the high rejection efficiency achieved by this cut, the residual contribution is still one order of magnitude higher that the required one \cite{Bandac:2020}. 
The rejection efficiency can be increased to 99.2\% with the help of the veto system tagging  muon-induced events as coincidences between two sectors. However, this is still not sufficient to reach the $\sim$10$^{-3}$~cnts/keV/kg/yr background level required for the experiment. The efficiency can be further increased to 99.7\% when any event in coincidence with the veto modules is rejected. Therefore, the veto system was configured to a single sector based muon-trigger to reach the CROSS background objective.

\section{Optimization of the CROSS muon veto system}
\label{sec:Veto_Optimization}

To reach the background index required for the CROSS experiment, it is essential to configure the veto system to reject all the events in thermal detectors that are in coincidence with any of the veto modules. However, the results from the simulation (see figure~\ref{fig:Muon_induced_events}) indicate that the muon energy distribution in the veto modules is inside the range of the natural radioactivity, which means that without any coincidence requirement between sectors the trigger rate and the dead time are expected to be high. To clarify this point, a measurement campaign was carried out to find the optimal energy thresholds that increase the muon rejection and keep the dead time relatively low. This study was carried out in the following steps:
\begin{enumerate}
    \item Calibration of the veto modules, obtaining the relation between the energy threshold and the value of the threshold level applied with the muon veto controller.
    \item Measurement of the rate of the veto modules at each energy threshold.
    \item MC simulations to calculate the background index and dead time for different trigger rates in the \emph{Lateral} and \emph{Bottom} veto modules.
    \item The search for the trigger rates (and threshold levels) of each module that optimize the sensitivity of the CROSS experiment.
\end{enumerate}
It should be noted that all SiPMs in this study were biased at 31~V (3~V of the SiPM overvoltage) to ensure homogeneity (in contrast to the spread from 30 V to 31 V used in the previous operation of the veto system investigated in section \ref{sec:MC_validation}).

\subsection{Calibration of veto modules}

A single module of each sector was calibrated using a $^{60}$Co source, placed in the center of the polystyrene bars. The Raspberry Pi controller measured the rate during 100~s at each of the 64 available threshold levels. Since these measured rates account for all events with an amplitude higher than the threshold level, the amplitude spectrum was obtained as the derivative of the rates measured at each threshold level. The same process was followed without the radioactive source to apply a background subtraction of the calibration spectrum.

\begin{figure}
\centering
\includegraphics[width=0.485\textwidth]{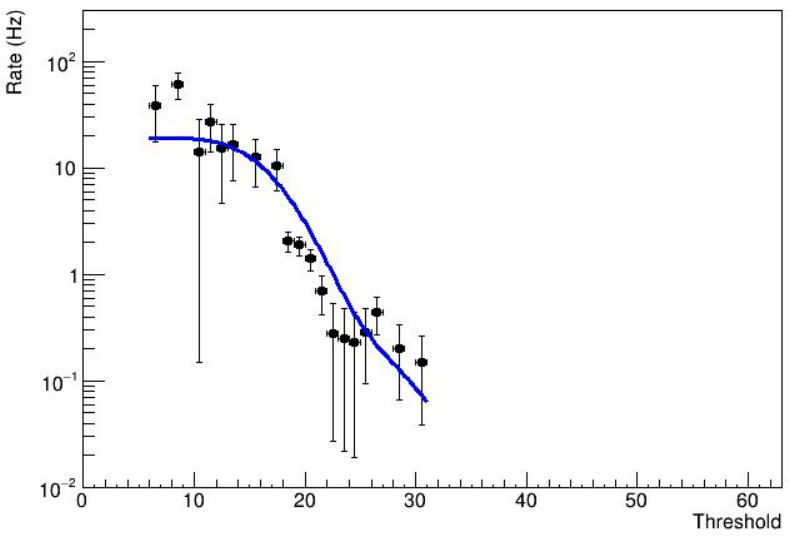}
\includegraphics[width=0.49\textwidth]{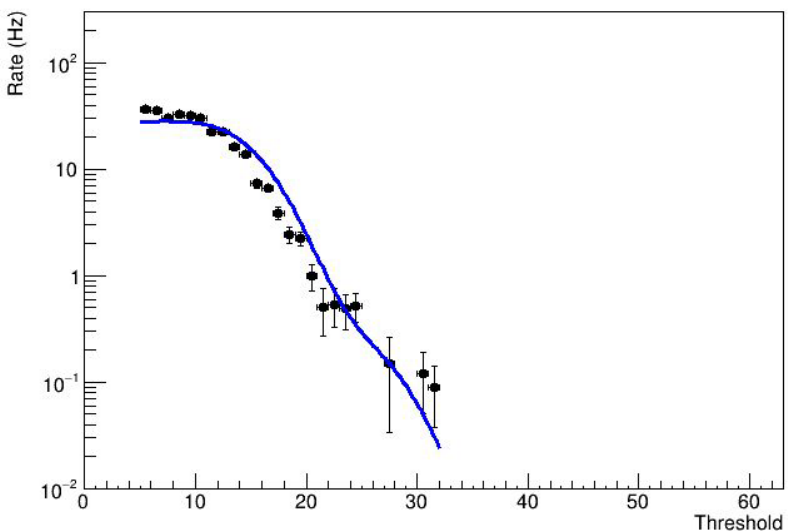}
\caption{Fit of the experimental spectra of pulse amplitudes measured in two veto modules irradiated by a $^{60}$Co source (after a background subtraction) to the simulated ones, which have been convolved in energy to account for the energy resolution of the detectors. The units of the amplitudes are given in threshold level units (37.5~mV each).
(Left) A \emph{Lateral} veto module of the S3 sector.
(Right) A \emph{Bottom} veto module of the S4 sector.
}
\label{fig:VetoCalibration}
\end{figure}

To establish a relation between the energy and the value of the threshold level, it was assumed that the amplitudes of SiPM pulses are proportional to their areas and that the area of the pulses is also proportional to the energy. Then a dedicated simulation of this calibration was run, and the obtained energy spectrum was convolved with a Gaussian function with constant $\sigma$ to account for the energy resolution of the detectors. Finally, the amplitude spectrum obtained in the calibration measurement was fitted to the simulated one assuming a proportional relation between the energy and the threshold level ($E=c\cdot th$). The free parameters of that fit were the calibration parameter ($c$) and the energy resolution value ($\sigma$). Figure~\ref{fig:VetoCalibration} shows examples of a fit to the calibration data of a single module of the \emph{Lateral} and \emph{Bottom} sectors.

\subsection{Characterization of background in veto modules}

\begin{figure}
\centering
\includegraphics[width=0.485\textwidth]{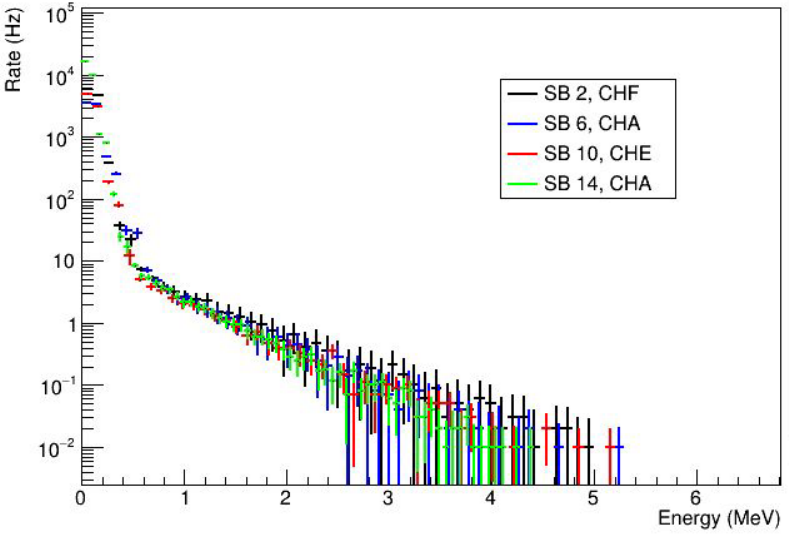}
\includegraphics[width=0.49\textwidth]{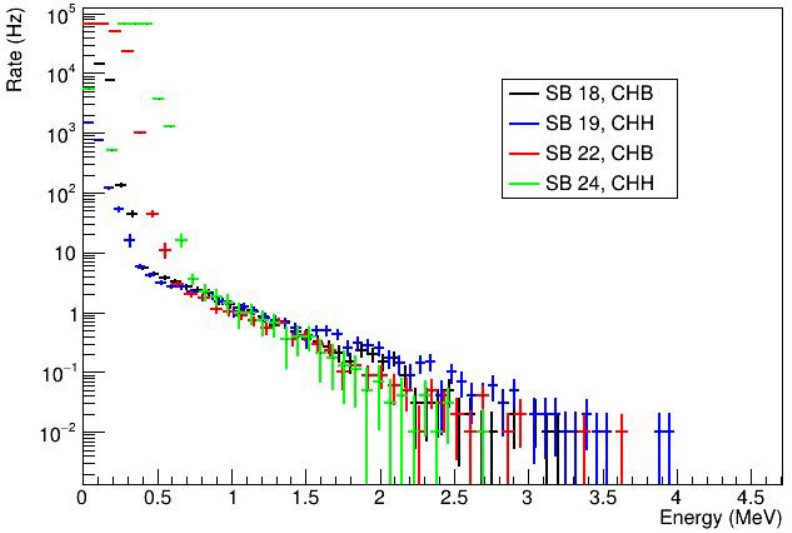}
\caption{Background rate calibrated in energy of four \emph{Lateral} (Left) and four \emph{Bottom} (Right) veto modules.}
\label{fig:VetoBackground}
\end{figure}

Once the relation between energy and threshold level was obtained, the background rates measured at each threshold level were translated into energy thresholds and compared to each other, as shown in figure~\ref{fig:VetoBackground}. This comparison shows that the rates for the same kind of veto module at each energy threshold are similar. This was expected since all modules are of the same size and the SiPMs are all biased to the same voltage. Moreover, the rate of the \emph{Bottom} modules is consistently lower than that of the \emph{Lateral} ones, which is explained by their reduced sizes.

There are two main components in the measured background: one with a high rate below $\sim$0.5~MeV and another one with lower rate at higher energies. The first one is likely due to noise events, while he other is identified as the natural radioactive background, which has an exponential decay with energy and is compatible with no rate at energies higher than $\sim$3~MeV. The second contribution to the measured energy-dependent rate $R(E)$ has been fitted to
\begin{equation}
    R(E) = a \cdot e^{-b \cdot E}
    \label{eq:VetoRateEnergy}
\end{equation}
to get the relation between the background rate at each energy threshold for each veto module. The fit was performed from 1~MeV to 2~MeV, in a range where the energy threshold should be set. The parameters $a$ and $b$ obtained in the fit are 13(2)~Hz and 1.7(1)~MeV$^{-1}$ for the \emph{Lateral} modules, and 19(5)~Hz and 2.6(1)~MeV$^{-1}$ for the \emph{Bottom} ones, respectively.

Thanks to this analysis, it was possible to directly relate the rate of a given module with the energy threshold applied. Therefore, to ensure homogeneity in the energy thresholds, the modules of the same part of the muon veto were configured to measure the same rate in all of them.

\subsection{Maximization of the CROSS experiment sensitivity}
\label{sec:optimization}

The next step in this analysis is to find the rates of the \emph{Lateral}, \emph{Bottom} and \emph{Top} modules that maximize the sensitivity of the CROSS experiment (expressed in half-life limit $T_{1/2}$ on $0\nu\beta\beta$ decay), which depends on the experiment's live time $t_{\textrm{Live}}$ and the background index $BI$ in the ROI as
\begin{equation}
T_{1/2}\propto\sqrt{t_{\textrm{Live}}/BI} .
\end{equation}
If we define the dead-time ratio  $r_{\textrm{Dead}}$ as the fraction of the total dead time $t_{\textrm{Dead}}$ over the total measurement time of the experiment $t_{\textrm{Total}}$, then the live time is
\begin{equation}
t_{\textrm{Live}} = t_{\textrm{Total}} \cdot (1-r_{\textrm{Dead}}) .
\end{equation}
If $t_{\textrm{Window}}$ is the time coincidence window applied between thermal detectors and muon veto modules, the dead time of the experiment is just 
\begin{equation}
r_{\textrm{Dead}} = 1-e^{-t_{\textrm{Window}} \cdot R_{\textrm{Total}}} ,
\end{equation}
where $R_{\textrm{Total}}$ is the sum of the rates of all veto modules. 
Since all modules of the same part of the muon veto are configured to have the same rate, the total muon veto rate $R_{\textrm{Total}}$ is given by
\begin{equation}
    R_{\textrm{Total}} = R_{\textrm{Lateral}} \cdot N_{\textrm{Lateral}} + R_{\textrm{Bottom}} \cdot N_{\textrm{Bottom}} +R_{\textrm{Top}} \cdot N_{\textrm{Top}},
\end{equation}
where $R_{\textrm{Lateral,Bottom,Top}}$ and $N_{\textrm{Lateral,Bottom,Top}}$ are the rate of each module and the number of modules in each part of the muon veto for the \emph{Lateral}, \emph{Bottom} and \emph{Top} sectors, respectively. However, since veto triggers are registered with the FPGA GPI/O channels of the CROSS DAQ, this rate is limited by the sampling rate $R_{\textrm{Sampling}}$ of the experiment, which is typically set at 2~kHz in the CROSS setup. It means that two consecutive triggers cannot be identified if they are produced within a time window of 0.5~ms and it reduces the total rate as
\begin{equation}
    R'_{\textrm{Total}} =R_{\textrm{Sampling}} \cdot \left(1 - e^{-R_{\textrm{Total}}/R_{\textrm{Sampling}}}\right).
\end{equation}

\begin{figure}
\centering
\includegraphics[width=0.65\textwidth]{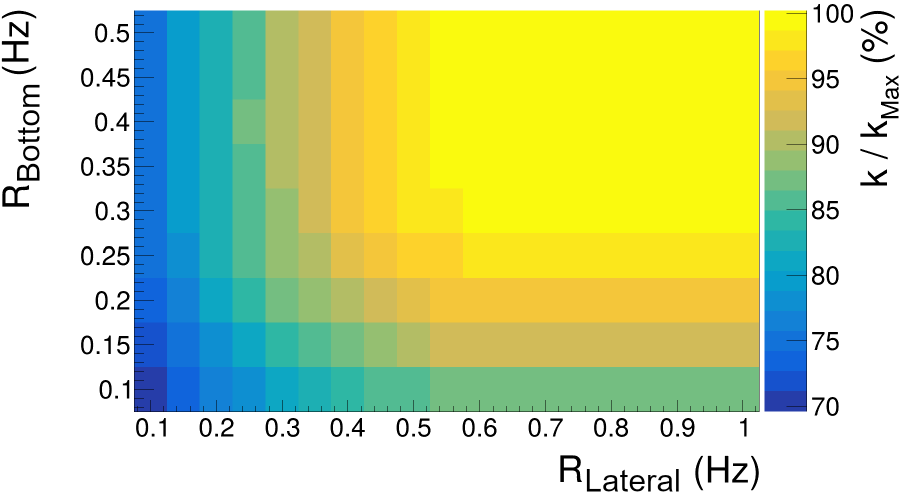}
\caption{Color-map of the $k$ parameter (proportional to the sensitivity of the CROSS experiment to $^{100}$Mo $0\nu\beta\beta$ decay), normalized to the maximum value, for different values of the \emph{Lateral} and \emph{Bottom} veto module rates.}
\label{fig:VetoBackground_LvsB}
\end{figure}

Thus, we can calculate the live time for a given set of rate values. Moreover, since the energy threshold applied to a given module is known through Eq.~\ref{eq:VetoRateEnergy}, it can be introduced into simulations to calculate the background index for the same set of rate values. To compare the sensitivities for each set of rate values, the parameter $k$ is defined as
\begin{equation}
    k=\sqrt{\frac{e^{-t_{\textrm{Window}} \cdot R'_{\textrm{Total}}}}{BI}},
\end{equation}
which is proportional to the CROSS sensitivity and only depends on the time coincidence window between the thermal detectors and the muon veto modules, the rate of each module and the background index. The rate of the \emph{Top} sectors was set to 0.6~Hz, since previous calibrations showed that it corresponds to an energy threshold of $\sim$1~MeV. Then, the $k$ value was calculated considering a time coincidence window of 2~ms, and the rates from 0.1~Hz to 1~Hz (0.5~Hz) for the \emph{Lateral} (\emph{Bottom}) modules, in steps of 0.05~Hz; the results are shown in figure~\ref{fig:VetoBackground_LvsB}.

This analysis shows that the sensitivity increases together with the rates of the modules, since it decreases the energy threshold. The sensitivity increases to a maximum value at $R_{\textrm{Lateral}}$ = 0.7~Hz and $R_{\textrm{Bottom}}$ = 0.3~Hz, slowly decreasing at higher rates due to the increase of the dead time. These rates correspond to energy thresholds of 1.7(1)~MeV and 1.6(1)~MeV for the \emph{Lateral} and \emph{Bottom} veto modules, respectively.

The threshold level of each veto module was set to obtain this trigger rate, resulting in a total rate of 19.6~Hz and 4.5~Hz for the \emph{Lateral} and \emph{Bottom} veto sectors, respectively. The total rate of the muon veto system, $R_{\textrm{Total}}$, was measured as 97.6~Hz, but the trigger registered with the CROSS DAQ, $R'_{\textrm{Total}}$ was a bit lower, 95.3~Hz, due to the sampling rate of 2~kHz. This corresponds to the dead-time ratio of 17.3\% if we set a time coincidence between the thermal detectors and the veto modules of 2~ms.

\section{Validation of the muon veto prior to the CROSS experiment}
\label{sec:Veto_Validation}

Before the installation of the CROSS demonstrator in the CROSS facility, we performed several tests of the veto system to ensure that it is possible to set the threshold level of all the veto modules, keeping the system in a stable mode with the slow control, and to identify coincident events between bolometers and the veto modules. This study was carried out in the cryogenic RUN~13, in which two CUPID-Mo modules (see figure~\ref{fig:MC_geometry2}) were installed and operated in the CROSS cryogenic facility.

\subsection{Operation of the muon veto system}

\begin{figure}
\centering
\includegraphics[width=0.45\textwidth]{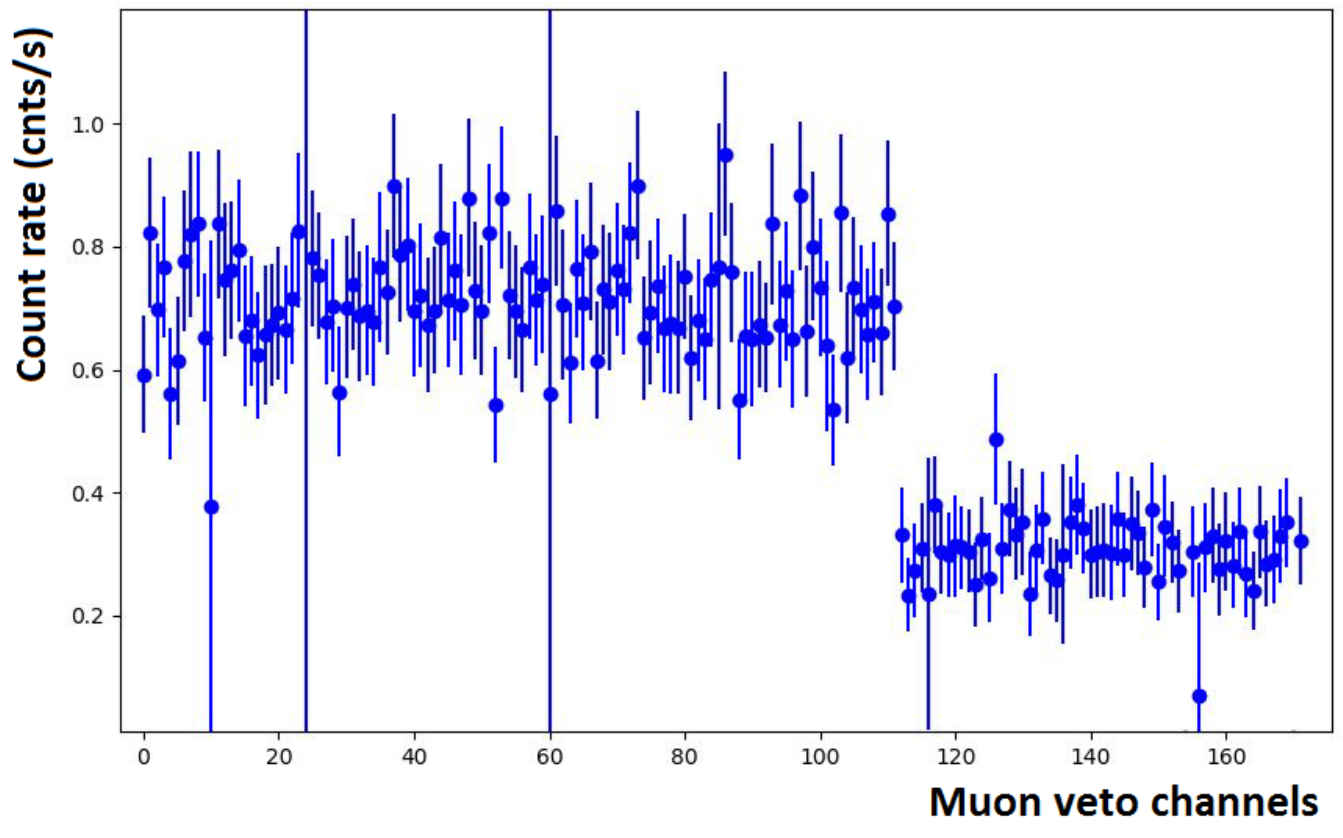}
\includegraphics[width=0.54\textwidth]{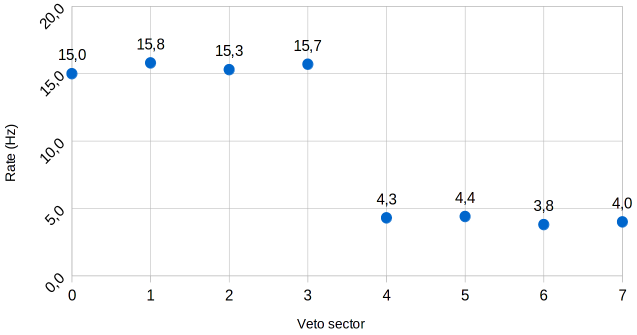}
\caption{Average rate of the CROSS muon veto;
Left) Measured with the Raspberry Pi controller in each channel of the \emph{Lateral} (first 112 channels) and \emph{Bottom} (113--172) channels of the CROSS muon veto;
(Right) Measured with the CROSS DAQ in each sector of the \emph{Lateral} (first four sectors) and \emph{Bottom} (last four sectors) parts of the CROSS muon veto.}
\label{fig:Veto_Rate_channels}
\end{figure}

The threshold level of the muon veto channels was configured using the Raspberry Pi controller to measure an average rate of 0.7~Hz in the \emph{Lateral} modules and 0.3~Hz in the \emph{Bottom} modules. The slow control continuously measured the rate of each individual channel for 60~s, while an acquisition with the CROSS DAQ was running, and no interference was observed between both. The rate measured with the slow control in each channel is shown in figure~\ref{fig:Veto_Rate_channels}, left; it demonstrates that most of the veto channels can be correctly set to get the required rate.

During this test, we discovered six SiPMs with unstable behavior, and it was not possible to stabilize them to get the correct trigger rate (the unstable SiPMs will be replaced soon). Apart from that issue, we also observed that the trigger bit provided by the veto trigger board and used in the CROSS DAQ to identify all triggers was deactivated. However, it was still possible to identify most of the triggers with the CROSS DAQ using the trigger pattern information provided by the other bits, since it updates when there is a trigger. The only triggers not identified are those whose trigger pattern is the same as the previous one. It implies a reduction in the total muon rate of 17\% (20\% in the \emph{Lateral} sectors and 4.6\% in the \emph{Bottom} ones). With this configuration, the total trigger registered in the CROSS DAQ should be 81.2~Hz, with 15.7~Hz and 4.3~Hz in the \emph{Lateral} and \emph{Bottom} sectors, respectively. 

The rate of each sector was also checked with the CROSS DAQ data. The average values are shown in figure~\ref{fig:Veto_Rate_channels}, right. All SiPMs of S1, S3, S4, and S5 were stable and operative, and therefore they have the expected rate calculated above. The other sectors had one or two non-biased SiPMs, and that is why the rate is a bit lower.

\begin{figure}
\centering
\includegraphics[width=0.7\textwidth]{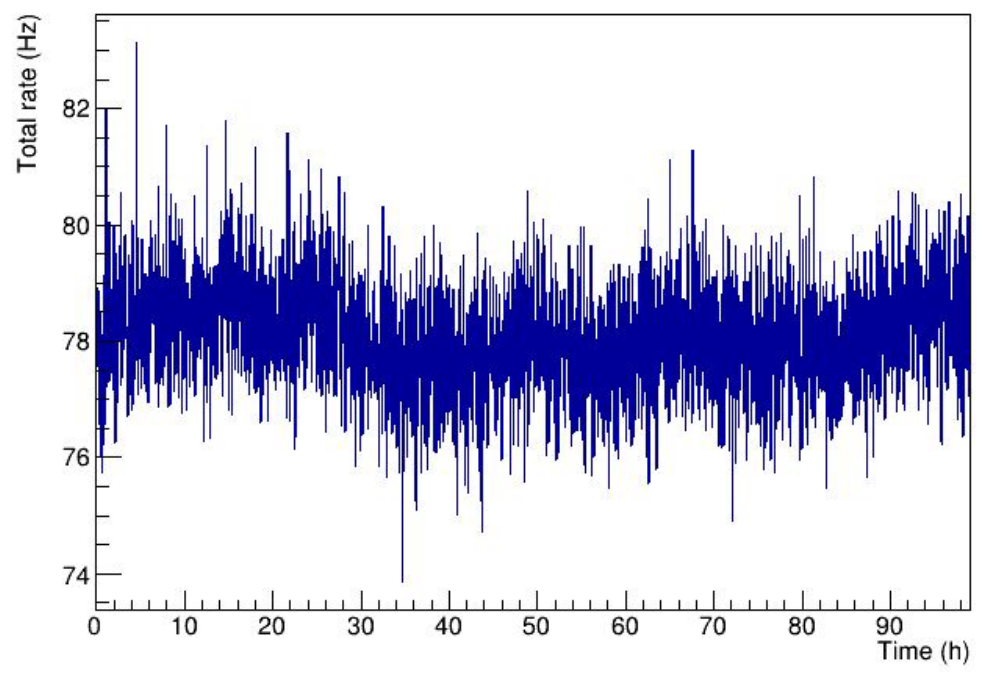}
\caption{Total trigger rate of the CROSS muon veto acquired with the CROSS DAQ during a background measurement over $\sim$100~h.}
\label{fig:Veto_stability}
\end{figure}  

The trigger bit was activated again a few weeks after the performance test presented here. Figure~\ref{fig:Veto_stability} shows the total trigger rate of the muon veto acquired with the CROSS DAQ during a background measurement over $\sim$100~h. Since the slow control was running during these measurements, the threshold level of each channel was corrected to keep the total rate as stable as possible. The rate instability during the acquisition time was found to be lower than 2\%, which is acceptable. The average rate was 78.3~Hz, a bit lower than the expected value of 81.2~Hz due to the six unstable SiPMs that were turned off during the measurement.

\subsection{Low-temperature measurement with two CUPID-Mo modules}
\label{sec:RUN13_test}

To validate the operation of the muon veto system and the cryogenic facility (after maintenance) prior to the CROSS experiment, we realized a low-temperature measurement (RUN~13) with two best-performance CUPID-Mo modules, LMO-5 and LMO-12 together with the corresponding light detectors LD-5 and LD-12 \cite{Armengaud:2020a}. In the framework of the CUPID-Mo experiment, these detectors were operated in the EDELWEISS setup \cite{Armengaud:2017b} at the Modane underground laboratory (LSM, France) to search for $0\nu\beta\beta$ decay \cite{Armengaud:2021,Augier:2022} and other rare-event processes \cite{Augier:2023,Augier:2024BSM}. Thus, investigation of the CUPID-Mo detector performance in the CROSS setup would provide a useful benchmark of the experimental conditions prior to the CROSS $0\nu\beta\beta$ decay search experiment.

Each CUPID-Mo module is made of a 0.21-kg LMO crystal ($\oslash$44 $\times$ 45 mm), produced from molybdenum enriched in $^{100}$Mo to around 97\%, and a Ge wafer ($\oslash$44 $\times$ 0.18 mm), assembled in a common Cu housing with PTFE supporting elements. Both absorber materials are instrumented with NTD Ge thermistors \cite{Haller:1994}; a 50-mg (10-mg) sensor is glued to the crystal (wafer) surface using Araldite\textregistered~ 2-component epoxy. A Si:P chip acting as a heater \cite{Andreotti:2012} is also glued on the LMO surface; it is used for constant power injection, allowing stabilization of the detector response with temperature fluctuations. To improve light collection efficiency, both Ge sides are covered with a 60-nm SiO antireflective coating, and the LMO crystal is surrounded by a reflective film. A module containing the LMO-5 crystal was placed on top of the LMO-12, thus its light detector LD-5 was able to detect scintillation of both crystals, reproducing the light collection conditions of these modules as in CUPID-Mo \cite{Armengaud:2020a}. 

\begin{figure}
\centering
\includegraphics[width=0.50\textwidth]{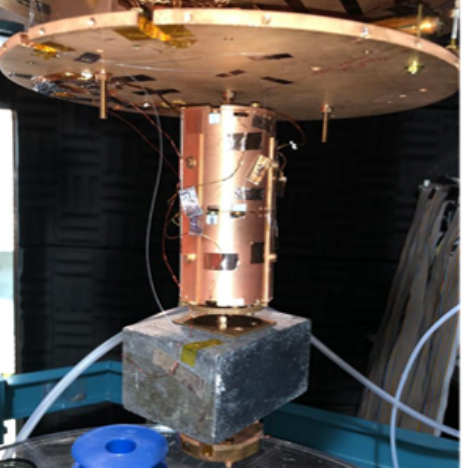}
\caption{Photo of a mini-tower made of two low-temperature detector modules with dual heat-scintillation readout, used in the CUPID-Mo experiment (LMO-5 and LMO-12 in \cite{Armengaud:2020a}), connected to the detector plate of the CROSS facility. A lead brick was added below the tower to increase the total weight, needed to optimize the detector suspension mitigating vibrational noise of the pulse-tube cryostat.}
\label{fig:r13_detectors}
\end{figure}  

The 2-crystal tower was mounted on the Cu detector plate of the CROSS cryogenic facility at the LSC (figure \ref{fig:r13_detectors}). This plate is suspended by three Kevlar ropes mechanically connected to magnetic dumpers installed at the 1-K stage of the dilution refrigerator \cite{CROSS_Magnetic_dampers:2023}. An extra weight (a lead brick) was added below the mini-tower to get a proper dumping, since the suspension system is adapted for the CROSS detector scale weight.

We cooled the modules down to 20 mK, as measured by a thermometer on the detector plate, and then stabilized the detector plate's temperature at that value via a PID control. This temperature is similar to the CUPID-Mo operation conditions (20.7 mK). The control and read out of the thermal detectors in the CROSS setup is done using room temperature electronics \cite{Carniti:2020,Carniti:2023}; a DAQ digitizes the voltage output of each sensor in continuous streams by a 24-bit ADC. All data in RUN 13 were acquired with a sampling rate of 2 kHz. An active Bessel filter is also used in the readout chain; the chosen cut-off frequency was 300 Hz. Data processing is performed with a Matlab-based tool \cite{Mancuso:2016}, which implements the optimum filtering technique \cite{Gatti:1986} using data-extracted information on average signal and noise. For each offline triggered event with a typical threshold of 5 rms of the baseline noise, the program computes the signal maximum (proportional to the deposited energy), the pre-trigger level (to monitor the temperature stability), and several pulse-shape parameters.

\begin{table}
\centering
\caption{Performance of two CUPID-Mo modules, based on thermal detectors made of a 0.2-kg Li$_2$$^{100}$MoO$_4$ crystal scintillator and a Ge wafer each, operated at 20 mK in the CROSS cryogenic facility (C2U) at the LSC. 
We report the NTD resistance $R_{NTD}$ of the working point, detector's sensitivity ($A_{signal}$), energy resolution (full width at half maximum) measured for zero energy deposition (FWHM$_{bsl}$) and for 2615 keV $\gamma$ of $^{208}$Tl (FWHM$_{2615}$). 
Statistical uncertainties of all values, except FWHM$_{2615}$, are below the given precision. 
Performance of these modules operated in the EDELWEISS (EDW) at 20.7 mK during the CUPID-Mo experiment \cite{Armengaud:2020a} and results of the best-performing CROSS module (LMO-best and LD-best), achieved in RUN~9 at 17~mK \cite{CROSS_Run9:2025}, are given for comparison. 
}
\smallskip
\begin{tabular}{ccccccc}
\hline
\multicolumn{2}{c}{Detector}  & Facility & $R_{NTD}$ & $A_{signal}$ & FWHM$_{bsl}$ & FWHM$_{2615}$  \\
ID & Design & ~ & (M$\Omega$) & (nV/keV) & (keV) & (keV)  \\
\hline
\hline
LMO-5    & CUPID-Mo & C2U & 4.4  & 39   & 3.4   & 6.7(5)  \\ %
~        & ~        & EDW & 1.8  & 21   & 1.5   & --  \\ %
LMO-12   & CUPID-Mo & C2U & 2.2  & 22   & 2.8   & 5.9(4)  \\ %
~        & ~        & EDW & 3.8  & 15   & 1.8   & --  \\ %
LMO-best & CROSS    & C2U & 4.8  & 84   & 1.1   & 5.7(3)  \\ %
\hline
LD-5     & CUPID-Mo & C2U & 3.2  & 2010 & 0.064 & --  \\ %
~        & ~        & EDW & 0.8  & 1100 & 0.066 & --  \\ %
LD-12    & CUPID-Mo & C2U & 1.7  & 1980 & 0.072 & --  \\ %
~        & ~        & EDW & 0.9  & 1000 & 0.069 & --  \\ %
LD-best  & CROSS    & C2U & 4.0  & 1830 & 0.056 & --  \\ %
\hline
\end{tabular}
\label{tab:LMO_performance}
\end{table}

\begin{figure}
\centering
\includegraphics[width=0.465\textwidth]{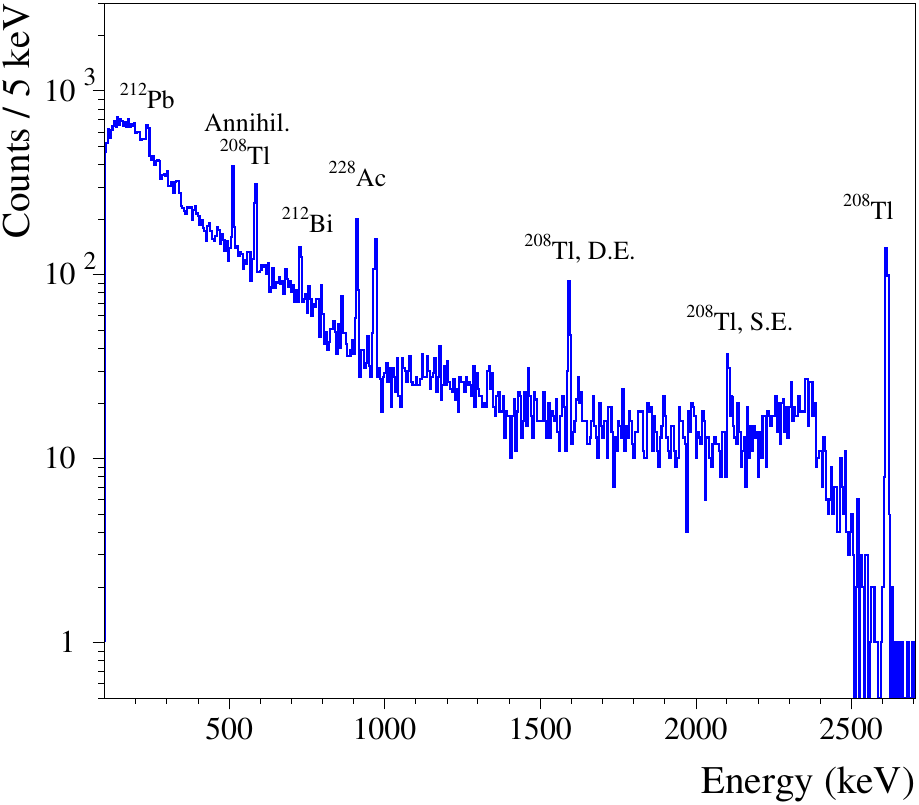}
\includegraphics[width=0.525\textwidth]{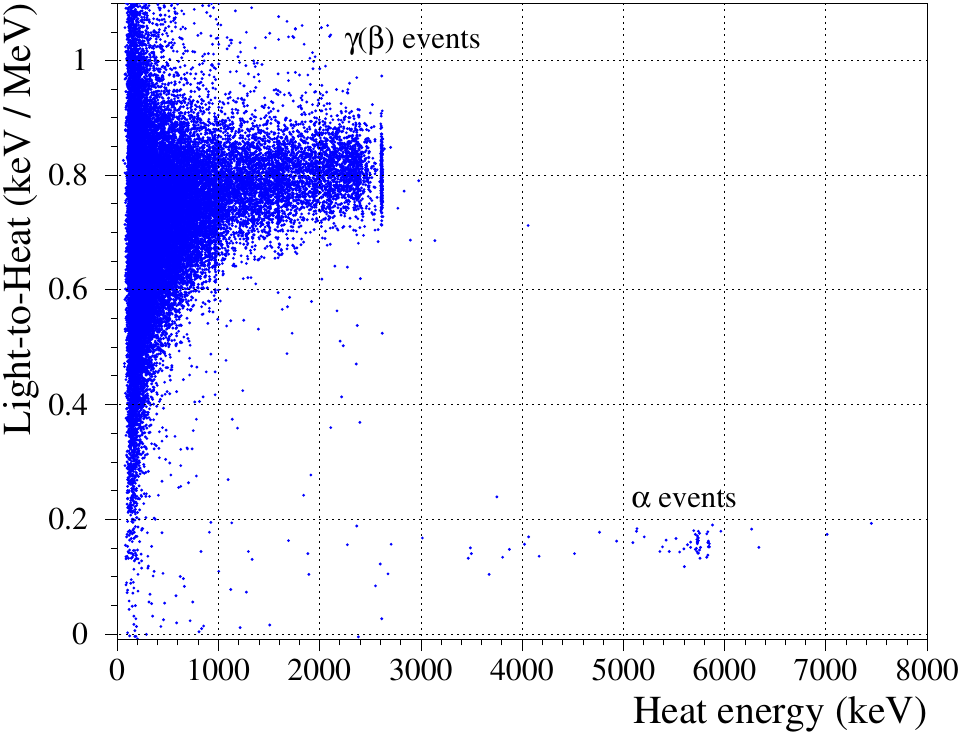}
\caption{(Left) Energy spectrum measured with a Li$_2$$^{100}$MoO$_4$ thermal detector (LMO-5) in 78-h-long calibrations with a $^{232}$Th source in the CROSS underground setup. The most intense $\gamma$-ray peaks found in the spectrum are labeled; D.E. and S.E. are double and single escape peaks, respectively. 
(Right) Distribution of the Light-to-Heat parameter versus heat energy of events detected by the LMO-5 module. 
}
\label{fig:r13_LMO_calibration}
\end{figure}  

Prior to data collection, we measured load curves --- V(I) characterization --- of the NTD-Ge thermistors aiming at selecting the working parameters above the inversion point. Thus, we set a 2~nA current across the thermistors, similarly to the CUPID-Mo working points, and measured a few M$\Omega$ resistances reported in Table \ref{tab:LMO_performance}. 

The CUPID-Mo modules were calibrated with a $^{232}$Th $\gamma$ source, which was inserted inside the lead shielding of the setup. The energy spectrum acquired by one LMO thermal detector (LMO-12) during a 78-h-long calibration measurement is shown in figure \ref{fig:r13_LMO_calibration}, left. Thanks to the determination of the energy scale, we measured the voltage signal output of LMO per unit of deposited energy (the so-called sensitivity) at a level of 20--40 nV/keV, a factor 1.5--2 higher than in the CUPID-Mo experiment. This can be explained by slightly colder temperature conditions in the present work. At the same time, fluctuations of the detectors' baseline ($\sim$3 keV FWHM) are found to be a factor 1.5--2 worse than the CUPID-Mo results. Nevertheless, both detectors demonstrate good energy resolution in a wide energy range, in particular $\sim$6--7 keV FWHM at 2615 keV $\gamma$ quanta of $^{208}$Tl, an energy of $\sim$400 keV less than the Q-value of $^{100}$Mo. These results are compatible with the best performance of the CROSS detectors operated in this setup, as presented in Table \ref{tab:LMO_performance}. The only evident difference in performance between the CUPID-Mo and CROSS LMO detectors is about a factor 4 higher sensitivity ($A_{signal}$) achieved because of the use of UV-cured glue to couple the sensor and it allowed us to get a lower baseline noise. 

The Ge light detectors were calibrated by 17.4-keV Mo X-rays, emitted by LMO crystals under the $^{232}$Th source-induced $\gamma$ irradiation. As a result of the colder working points (a factor 2--3 higher resistances), both devices are characterized by a similar sensitivity of 2 $\mu$V/keV, a twice higher value than in CUPID-Mo. However, the baseline noise achieved, $\sim$60--70 eV FWHM, is very close to the CUPID-Mo performance and to the best results obtained with a CROSS LD (see Table \ref{tab:LMO_performance}). 

The normalization of Ge LD signals on the heat energy release measured in coincidences by an accompanied LMO, the so-called Light-to-Heat ratio, is often used as a particle identification parameter. This is illustrated in figure \ref{fig:r13_LMO_calibration} (right), where the band of $\gamma$($\beta$) events is clearly separated from the distribution of $\alpha$ particles, characterized by a quenched scintillation emission (around 20\% for equivalent energy $\gamma$($\beta$)s). The results of the mean Light-to-Heat ratio for high-energy, 2--3 MeV,  $\gamma$($\beta$) events are reported in Table \ref{tab:LMO_LY}; the obtained results are in good agreement with the CUPID-Mo data \cite{Armengaud:2020a}.

\begin{table}
\centering
\caption{Mean values of the Light-to-Heat parameter for the $\gamma$($\beta$) events with energies 2--3 MeV measured by two CUPID-Mo modules operated in the C2U setup at the LSC. The LMO-12 crystal was viewed by a cryogenic light detectors from both sides, while only a single photodetector was coupled to the LMO-5 sample. This configuration was also used in CUPID-Mo \cite{Armengaud:2020a}, the results of which (obtained in the EDW setup) are given for comparison. An RMS of the the Light-to-Heat distribution is taken as a conservative estimate of the mean value uncertainty.
}
\smallskip
\begin{tabular}{cccc}
\hline
Detector & Facility & \multicolumn{2}{c}{Light-to-Heat ratio (keV/MeV)} for $\gamma$($\beta$) events \\
ID &  ~ & Bottom LD & Top LD  \\
\hline
\hline
LMO-5    & C2U & 0.81(4)  & --  \\ %
~        & EDW & 0.80  & --  \\ %
LMO-12   & C2U & 0.64(4)  & 0.71(4)     \\ %
~        & EDW & 0.65  & 0.75    \\ %
\hline
\end{tabular}
\label{tab:LMO_LY}
\end{table}

After the $^{232}$Th calibration that allowed us to understand the detector performance, we acquired a long dataset (555 h) without sources aiming to investigate the efficiency of coincidences with the muon veto to reject events around $^{100}$Mo $0\nu\beta\beta$ ROI. The results of this study are detailed in the next section.

\subsection{Coincidences between thermal detectors and muon veto}
\label{sec:Veto_efficiency}

It is essential to keep the time coincidence window between thermal detectors and the veto modules as small as possible. Therefore, the trigger positions of the LD signals, which are much faster than the corresponding heat signals of the crystals, were used to identify coincidences with the veto modules. The position of a light signal was obtained as a corresponding heat trigger position (set at the signal's maximum) minus the offset between them to account the time difference in the rising part of the thermal signals (as used in figure \ref{fig:r13_LMO_calibration}, right). 
For each event identified in a crystal-based thermal detector, the minimum time difference between a light signal and a muon veto trigger, $\Delta t$, was obtained.

\begin{figure}
\centering
\includegraphics[width=0.7\textwidth]{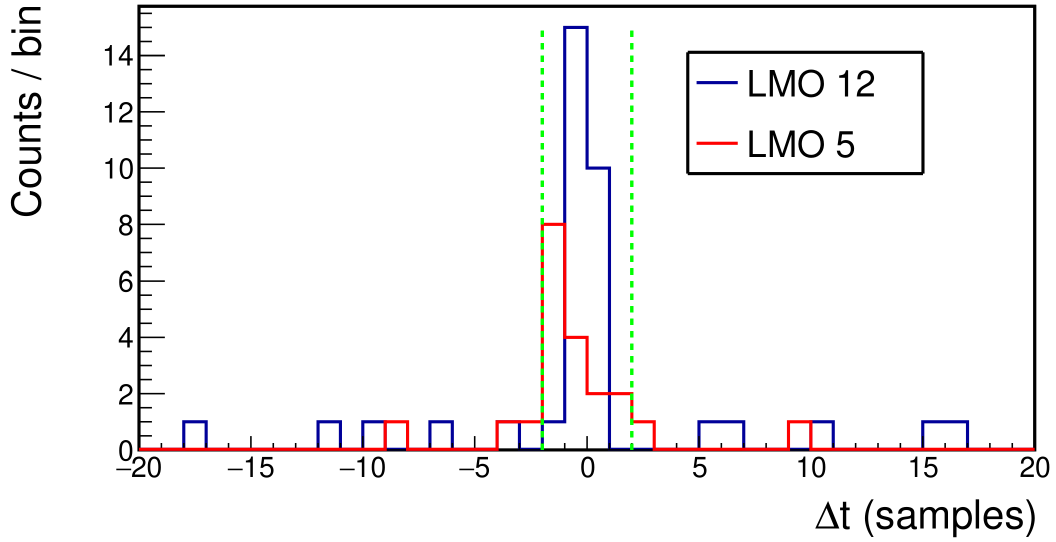}
\caption{Coincidences between two LMO thermal detectors and the muon veto measured in the CROSS cryogenic setup (555 h of data) at the LSC. The event selection is the following: Energy (heat) > 3 MeV and the Light-to-Heat value corresponding to the $\gamma$($\beta$) band. Green dashed lines represent the coincidence window ($\pm$1~ms) used in this analysis.}
\label{fig:r13_LMO_Veto_coincidences}
\end{figure}

Figure~\ref{fig:r13_LMO_Veto_coincidences} represents the variable $\Delta t$ for muon-like events in the LMO thermal detectors of the two CUPID-Mo modules. We selected events with an energy deposit in the crystal higher than 3~MeV and the Light-to-Heat value corresponding to the $\gamma$($\beta$) band (selected with a window of 3$\sigma$ around the mean value listed in Table \ref{tab:LMO_LY}). 
This figure clearly shows a high contribution of time-correlated events between veto modules and both LMOs. By opening a coincidence time window of $\pm$1~ms (4 samples), the rejection power was 70\% in both LMOs, with a dead-time ratio of 15\%. These parameters should be increased to 84\% and 18\%, respectively, since in these measurements we faced a 17\% reduction in the muon veto trigger rate due to the disabled ``trigger bit''.

\begin{figure}
\centering
\includegraphics[width=1.00\textwidth]{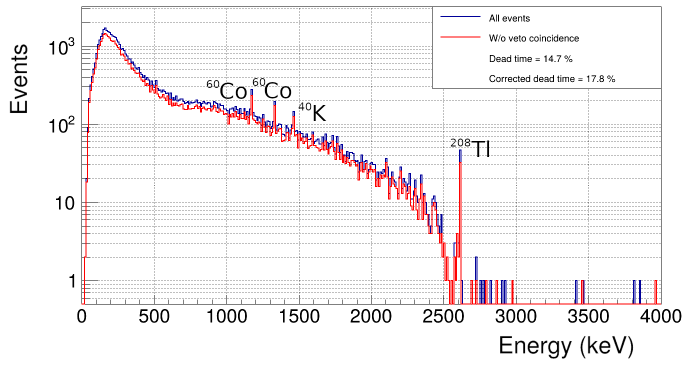}
\caption{Energy spectrum of $\gamma$($\beta$) events detected in the 555-h-long background measurements with two thermal detectors based on 0.2 kg $^{100}$Mo-enriched LMO crystal scintillators (previously used in CUPID-Mo) operated in the CROSS cryogenic facility at the LSC. The acquired events are selected in the Light-to-Heat interval corresponding to the $\gamma$($\beta$) band (blue histogram). 
The impact of the muon veto system on the background reduction is demonstrated by the anticoincidence cut applied (red histogram), allowing to remove events in the region of interest for $^{100}$Mo $0\nu\beta\beta$ decay search (around 3 MeV). }
\label{fig:r13_LMO_Bkg}
\end{figure}  

Figure~\ref{fig:r13_LMO_Bkg} shows the spectrum of $\gamma$($\beta$) events measured by the two LMO thermal detectors over 555 h of data taking; the events are selected in the $\pm$3$\sigma$ interval of the Light-to-Heat values for $\gamma$($\beta$)s and requiring the detector multiplicity 1 (an energy deposition is in a single detector). 
The resulting energy spectrum has several prominent features:
\begin{itemize}
    \item an exponential rise of events at energies below $\sim$700 keV is mainly induced by the bremsstrahlung $\gamma$ radiation emitted in the $^{210}$Bi decay, a daughter of  $^{210}$Pb present in the lead shield \cite{CROSSdeplLMO:2023};
    
    \item a background continuum in 700--2500 keV range is dominated by $^{100}$Mo two-neutrino double-beta decay; since we operated the $^{100}$Mo-enriched detectors, the counting rate of this process is expected to be on the level of 2 mHz \cite{Armengaud:2020a,Augier:2023model};
    
    \item a doublet of $^{60}$Co $\gamma$ quanta (1173 and 1333 keV) from a surface contamination of the LD-5, as observed in the CUPID-Mo experiment \cite{Armengaud:2020a,Augier:2023model}; 
    
    \item $\gamma$ peaks of $^{208}$Tl (2615 keV) and $^{40}$K (1461 keV) from the residual radioactivity inside the cryostat \cite{CROSSdeplLMO:2023};
    
    \item muon-induced events with energies above the end-point (i.e., 2615-keV $\gamma$ quanta of $^{208}$Tl) of the most intense natural radioactivity.
\end{itemize}

The impact of the muon veto system is clearly observed in the reduction of events in the region of interest for the search for $^{100}$Mo $0\nu\beta\beta$ decay, as shown in figure \ref{fig:r13_LMO_Bkg}. A muon simulation with the RUN~13 geometry configuration was performed to verify the efficiency of the muon veto system. Table~\ref{tab:MuonRejection} shows the muon veto efficiency to reject muon-like events with energies greater than 3 or 10~MeV. The values of this efficiency after the reset of the ``trigger bit'' are also reported together with those obtained with the simulation. We found a good agreement between the experimental and simulation results, meaning that we can expect a reduction of $\sim$80\% of muons in the CROSS experiment that was reported above by the simulation of this configuration (see section \ref{sec:optimization}).

\begin{table}
    \caption{Percentage of muon-induced events with the energy deposition above 3 or 10 MeV in any of two LMOs (CUPID-Mo modules), which are detected in coincidences with the muon veto according to the experimental data and MC simulations. We also quote the ``corrected'' experimental results, taking into account a 17\% increase in the muon veto trigger rate once the ``trigger bit'' is restarted (see details in text).}
    \centering
    \begin{tabular}{c|ccc}
    \hline
        Event  & \multicolumn{3}{c}{Muon-induced events of two LMOs in coincidences with any veto sector} \\ 
        \cline{2-4}
        energy & Experiment & Experiment (corrected) & Simulation \\ 
        \hline
        $E$ > 3~MeV & 71\% & 85\% & 88\% \\
        $E$ > 10~MeV & 75\% & 90\% & 89\% \\
        \hline
    \end{tabular}
    \label{tab:MuonRejection}
\end{table}

\section{Conclusions}

The CROSS (Cryogenic Rare event Observatory with Surface Sensitivity) Collaboration is preparing a high-sensitivity searches for neutrinoless double-beta decay of $^{100}$Mo with low-temperature detectors to be operated in a dedicated low-background setup at the Canfranc underground laboratory (LSC) in Spain. Given a comparatively high residual flux of cosmic muons ($\sim$20 $\mu$/m$^2$/h) in the LSC, we designed and installed a muon veto system around the CROSS facility. The muon veto is based on polystyrene bars and divided on nine sectors: 4 lateral, 1 top, and 4 bottom sectors. The lateral and bottom modules are readout with Si photomultipliers (SiPMs), while the top sector uses two conventional PMTs in OR mode. 

To investigate the correctness of the veto system operation and to optimize it, we realized Monte Carlo (MC) simulations of muons passing through the CROSS setup with a detector array of three geometries: 
(a) a 10-crystal tower with 45-mm-side cubic crystals, six of them are based on lithium molydbate (LMO) and four samples are tellurium dioxide (TeO), a CROSS-like detector array tested in the facility \cite{CROSS_enriched_TeO:2024,CROSS_Run9:2025};
(b) a tower of two CUPID-Mo modules with cylindrical LMO crystals ($\oslash$44 $\times$ 45 mm each) \cite{Armengaud:2020a}, operated prior the CROSS experiment; 
(c) a 42-crystal tower (36 LMOs and 6 TeOs) of the CROSS experiment \cite{Bandac:2020}, which is going to start commissioning by the end of 2025.
Each crystal in the considered geometries was accompanied by a cryogenic light detector. 
In particular, the MC shows that the initial muon trigger pattern, two veto sectors, is not sufficient for the CROSS experiment, and thus the single sector trigger is adopted. This trigger logic results to a comparatively high dead time ($\sim$18\%) expected in the CROSS experiment, but it allows to reject around 99.7\% of muon-induced events to the region of interest (around 3 MeV), leading to the residual contribution to the background index on the level of $\sim$2 $\times 10^{-3}$ cnts/keV/kg/yr. 
This configuration was tested in the last cryogenic run of the CROSS setup prior to the CROSS experiment. We demonstrate a stable operation of the veto system and a highly-efficient rejection of muon-induced events in the region of interest thanks to coincidences of LMO thermal detectors with any sector of the muon veto, as confirmed by dedicated MC simulations.

\acknowledgments

This work is supported by the European Commission (Project CROSS, Grant No. ERC-2016-ADG, ID 742345), by the Agence Nationale de la Recherche (ANR France; Project CUPID-1, ANR-21-CE31-0014), and US National Science Foundation (NS~1614611). We acknowledge also the support of the P2IO LabEx (ANR-10-LABX0038) in the framework ``Investissements d'Avenir'' (ANR-11-IDEX-0003-01 -- Project ``BSM-nu'') managed by ANR, France. 
This work was also supported by the National Academy of Sciences of Ukraine in the framework of the project ``Development of bolometric experiments for the search for double beta decay'', the grant number 0121U111684. 
Russian and Ukrainian scientists have given and give crucial contributions to CROSS. For this reason, the CROSS collaboration is particularly sensitive to the current situation in Ukraine. The position of the collaboration leadership on this matter, approved by majority, is expressed at \href{https://a2c.ijclab.in2p3.fr/en/a2c-home-en/assd-home-en/assd-cross/}{https://a2c.ijclab.in2p3.fr/en/a2c-home-en/assd-home-en/assd-cross/}.

\bibliographystyle{JHEP}

\end{document}